\documentclass[
showpacs,
floatfix,
aps,
prl,
twocolumn,
superscriptaddress,
amssymb,
]{revtex4-1}

\usepackage{amssymb}
\usepackage{latexsym}
\usepackage[normalem]{ulem}
\usepackage{graphicx}
\usepackage{amsmath}
\usepackage{dcolumn}
\usepackage{amsfonts}
\usepackage{bm}
\usepackage{color}
\usepackage{cleveref}

\newcommand{\rref}[1]{Ref.\,\onlinecite{#1}}

\newcommand{\be}{\begin{equation}}
\newcommand{\ee}{\end{equation}}
\newcommand{\bea}{\begin{eqnarray}}
\newcommand{\eea}{\end{eqnarray}}

\def\lb{\left [}
\def\rb{\right ]}

\def\wse2{WSe$_2$}
\def\dsas{\Delta_{\text{SAS}}}
\def\nutot{\nu_{\text{tot}}}

\newcommand{\req}[1]{Eq.\,(\ref{#1})}

\newcommand{\rfig}[1]{Fig.\,\ref{#1}}

\begin{document}

\title{Bilayer WSe$_2$ as a natural platform for interlayer exciton condensates in the strong coupling limit}

\author{Qianhui Shi}
\affiliation{Department of Physics, Columbia University, New York, NY, USA}
\author{En-Min Shih}
\affiliation{Department of Physics, Columbia University, New York, NY, USA}
\author{Daniel Rhodes}
\affiliation{Department of Mechanical Engineering, Columbia University, New York, NY, USA}
\author{Bumho Kim}
\affiliation{Department of Mechanical Engineering, Columbia University, New York, NY, USA}
\author{Katayun Barmak}
\affiliation{Department of Applied Physics and Applied Mathematics, Columbia University, New York, NY, USA}
\author{Kenji Watanabe}
\affiliation{Research Center for Functional Materials,
National Institute for Materials Science, 1-1 Namiki, Tsukuba 305-0044, Japan}
\author{Takashi Taniguchi}
\affiliation{International Center for Materials Nanoarchitectonics,
National Institute for Materials Science,  1-1 Namiki, Tsukuba 305-0044, Japan}
\author{Zlatko Papi\'c}
\affiliation{School of Physics and Astronomy, University of Leeds, Leeds LS2 9JT, UK}
\author{Dmitry A. Abanin}
\affiliation{Department of Theoretical Physics, University of Geneva,24 quai Ernest-Ansermet, 1211 Geneva, Switzerland}
\author{James Hone}
\affiliation{Department of Mechanical Engineering, Columbia University, New York, NY, USA}
\author{Cory R. Dean}
\affiliation{Department of Physics, Columbia University, New York, NY, USA}

\date{\today}

\maketitle

\textbf{Exciton condensates (EC) are macroscopic coherent states arising from condensation of electron-hole pairs \cite{blatt:1962}. Bilayer heterostructures, consisting of two-dimensional electron and hole layers separated by a tunnel barrier, provide a versatile platform to realize and study EC \cite{lozovik:1975b,snoke:2002,eisenstein:2004}. The tunnel barrier suppresses recombination yielding long-lived excitons \cite{butov:2002,eisenstein:2014,liu:2017,li:2017a,wang:2019,ma:2021}. However, this separation also reduces interlayer Coulomb interactions, limiting the exciton binding strength. Here, we report the observation of EC in naturally occurring 2H-stacked bilayer \wse2. 
In this system, the intrinsic spin-valley structure suppresses interlayer tunneling even when the separation is reduced to the atomic limit, providing access to a previously unattainable regime of strong interlayer coupling. 
Using capacitance spectroscopy, we investigate magneto-EC, formed when partially filled Landau levels (LL) couple between the layers. 
We find that the strong-coupling EC show dramatically different behaviour compared with previous reports, including an unanticipated variation of EC robustness with the orbital number, and find evidence for a transition between two types of low-energy charged excitations. 
Our results provide a demonstration of tuning EC properties by varying the constituent single-particle wavefunctions.
}

Narrow-gap semiconductors or semi-metals were originally proposed as natural hosts for EC due to a spontaneous paring instability between the electrons and holes, provided that the Coulomb interaction between them exceeds the band gap~\cite{halperin:1968}.
Signatures of EC have been identified in  several such candidate systems \cite{lu:2017,kogar:2017,zli:2019,kim:2020,jia:2020,ataeie:2021}, though debates are still ongoing due to challenges to disentangle lattice-related effects \cite{baldini:2020,kim:2020,watson:2020,mazza:2020}. 
In another direction, EC have been explored in layered structures where electrons and holes are spatially confined to separate layers \cite{lozovik:1975,butov:2004,min:2008,fogler:2014}.
These so-called spatially indirect excitons can be generated by either optical excitation or electrical gating \cite{snoke:2002,jauregui:2019,ma:2021}.
The spatial separation of the electrons and holes inhibits recombination and therefore extends the exciton lifetime, making it possible to observe the EC under equilibrium conditions \cite{ma:2021}. 

Quantum Hall bilayers provide a robust model platform to study the spatially indirect EC~\cite{eisenstein:2004,eisenstein:2014,li:2017a,liu:2017}. 
In these systems, a heterostructure consisting of electrically-isolated parallel layers is exposed to a perpendicular magnetic field.  Interlayer excitons can then form between partially filled LLs with  filled (electron) states in one layer coupling to vacancy (hole) states in the other layer. A key advantage of this approach is that, within the flat LLs, kinetic energy is quenched and Coulomb interactions necessarily play a dominant role. Quantum Hall bilayers fabricated from GaAs double wells  provided the first direct evidence of exciton superfluidity~\cite{nandi:2012,eisenstein:2014}, appearing when each layer was tuned to half filling of the lowest Landau level.  More recently, superfluid EC have been studied in graphene heterostructures, consisting of two graphene layers separated by a boron nitride tunnel barrier, where increased flexibility in device architecture has expanded the accessible phase space~\cite{liu:2017,li:2017a}.

\begin{figure*}[htb]
\begin{center}
\includegraphics{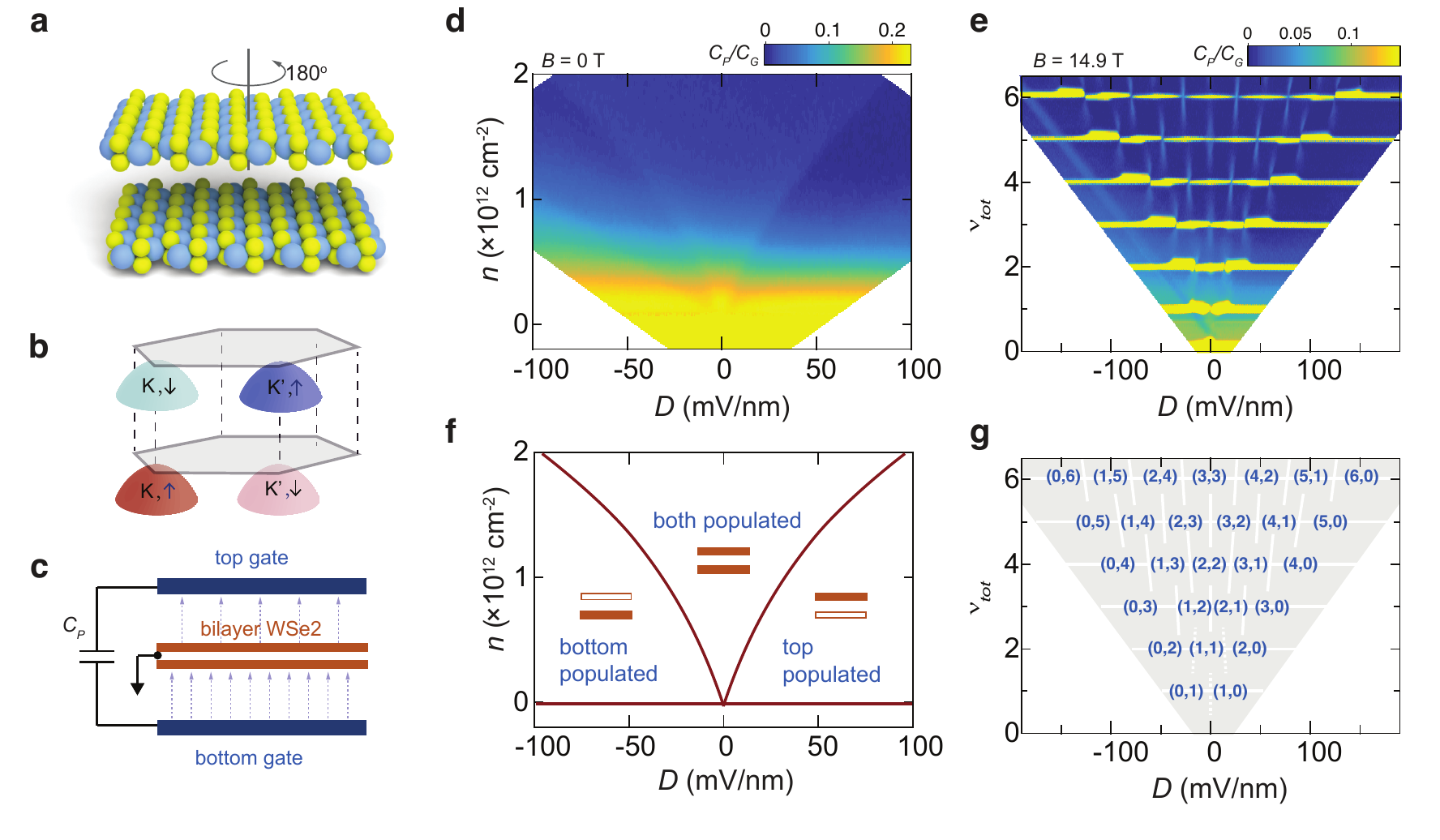}
\vspace{-0.15 in}
\end{center}
\caption{\small{
{\bf{Independent layer population in bilayer \wse2.}}
(a) Schematic of \wse2 bilayer crystal structure.
(b) Schematic illustration of the four flavors in the valence band in bilayer \wse2.
(c) Schematic illustration of the penetration capacitance measurements.
(d) Penetration capacitance of bilayer \wse2 measured at $B = 0$ and $T = 0.3$ K, versus $\nu_{tot}$ and $D$.
(e) Schematic phase diagram of layer population according to features in (c).
(f) Penetration capacitance measured at $B = 14.9$ T and $T = 0.3$ K, versus $\nu_{tot}$ and $D$.
(g) Schematic illustration of features shown in (e). The filling factors in each layers are marked as $(\nu_T,\nu_B)$ for the integer gaps.
}
}
\label{fig1}
\vspace{-0.15 in}
\end{figure*}

The EC phase diagram is determined by the interplay between the interlayer Coulomb attraction, $E_{inter}\propto1/d$, where $d$ is the layer separation,  and intralayer Coulomb repulsion, $E_{intra}\propto 1/l_{B}$, where $l_{B}=\sqrt{\hbar/eB}$ is the magnetic length \cite{eisenstein:2014}. Whereas $l_{B}$ can be widely varied with $B$, $d$ is more restricted since it must be small enough to promote strong coupling between the layers, at the same time remaining large enough to suppress interlayer tunneling. In GaAs bilayers,  $d>10$~nm, and the  EC are measurable only in the so-called weak coupling limit ($d/l_{B}\sim1$) \cite{eisenstein:2014}. Recent efforts in graphene double layers demonstrated the ability to access  the strong coupling regime ($d/l_{B}<0.5$), by reducing the layer separation to just a few nanometers \cite{liu:2020}. 
Owing to the inability to further reduce $d$ without introducing appreciable tunneling, the regime of extreme strong coupling remains almost completely unexplored \cite{kim:2021}.

\begin{figure*}[htb]
\begin{center}
\includegraphics{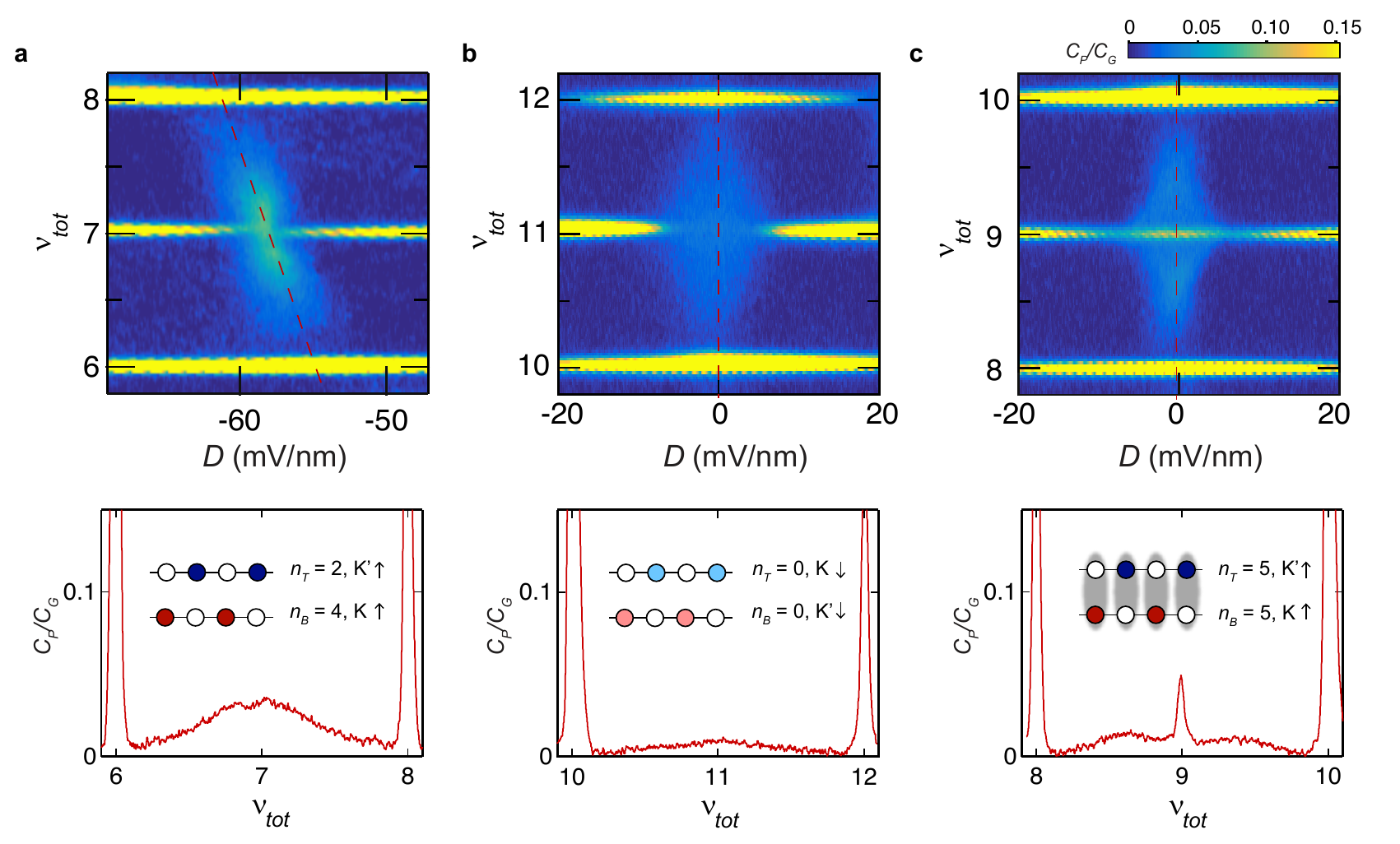}
\vspace{-0.15 in}
\end{center}
\caption{\small{
{\bf{EC formation and gap opening at LL crossings.}}
(a-c) Penetration capacitance versus total filling factor $\nu_{tot}$ and displacement field $D$ at $B = 14.9$ T around several LL crossings.
The bottom panel are linecut along the red dashed line in the top panel, which tracks equal topmost LL population in the two layers.
The insets of the bottom panel illustrate the orbital, spin and valley indices.
}}
\label{fig2}
\vspace{-0.15 in}
\end{figure*}   
 
Here we identify the natural bilayer \wse2 as a system that provides a unique opportunity to realize  EC in the previously inaccessible regime of extreme strong interlayer coupling.  In this case, interlayer tunneling is naturally suppressed via a spin-blocking mechanism, due to combination of the spin-valley locking and stacking order. This eliminates the need to insert a physical tunnel barrier, thereby allowing the layer separation to be reduced to the atomic limit.
Previously, in the weak coupling regime, EC has only been observed in the lowest LL ($n=0$). Remarkably, here we observe EC in higher LLs up to $n=6$.
Our data indicates that in the strong coupling regime the low-energy charged excitations have a different nature in different LLs, rendering the EC more robust in high LLs than in the lowest LL.

Fig. 1(a) shows a schematic of the crystal structure for 2H-stacked bilayer \wse2, consisting of two monolayers rotated 180 degrees to each other. 
In Fig.1(b) we sketch the low-energy valence bands, with the valley and spin indices indicated.  
Strong spin-orbit coupling gives rise to spin-valley locking within each layer, with the K and K' valleys oppositely spin-polarized. Due to the stacking order, carriers residing in the same valley have opposite spin in the two layers, leading to strongly suppressed interlayer tunneling \citep{gong:2013,zeng:2013,xu:2014,fallahazad:2016,pisoni:2019,pisoni:2019b} (also see SI.1).

In our experiment, we measure the penetration capacitance $C_P$, illustrated in Fig.1(c), which has proven to be an effective probe that circumvents non-ideal electrical contact and disorder effects in 2D semiconducting transition metal dichalcogenides \cite{shi:2020}.
While $C_P$ reflects the inverse compressibility of a monolayer system, for a multi-layer system it has an additional contribution from the polarizability.
The penetration capacitance normalized to the geometric capacitance between the top and bottom gates ($C_G$) can be written as \cite{young:2011,zibrov:2017},
\be
\frac{C_P}{C_G} = \frac{2 c \partial \mu/\partial n}{e^2+2 c \partial \mu/\partial n}+\frac{e c}{4 \epsilon_0 c_0}\frac{\partial \Delta n}{\partial D}.
\ee
Here, $c$ is the capacitance (per unit area) between the bilayer \wse2 and the gates, $c_0$ is the interlayer capacitance (per unit area) of the bilayer \wse2, $\partial\mu/\partial n$ is the inverse electronic compressibility where $\mu$ is the chemical potential and $n$ is the total carrier density, $\Delta n = n_T-n_B$ is the carrier density imbalance of the two layers, and $D = c(V_B-V_T)/2 \epsilon_0$ is the displacement electric field on the bilayer. 
The two terms in the right hand side of Eq.~(1) correspond to ``incompressibility'' and ``polarizability'' contributions, respectively; a large signal in $C_P$ indicates that the system is either incompressible or highly polarizable.

We first demonstrate that we can achieve layer-selective population in bilayer \wse2 through gate control.
In Fig.1(d) we plot the penetration capacitance versus the displacement field and total density. 
For a fixed carrier density, the electron polarization can be tuned through three regions, corresponding to the bottom, top or both layer populated.
The transition from full layer polarization to partial polarization is visible as a step in the $C_P/C_G$, owing to a finite polarizability contribution in the latter.
This step follows a curved shape that defines the polarization boundaries in the space of $D$ vs $n$ (see SI.3). 

The layer-selective population becomes more apparent as a perpendicular magnetic field is applied, and the energy spectrum splits into discrete LLs.
In Fig.1(e) we plot $C_P/C_G$ as a function of $D$ and total filling factor $\nu_{tot} = \nu_T + \nu_B$, where $\nu_T$ and $\nu_B$ are the top and bottom layer LL filling fractions, respectively. 
Horizontal features are observed at integer values of $\nu_{tot}$. These peaks in $C_P$ indicate incompressibility when the Fermi level is within an integer quantum Hall (IQH) gap.  In the bilayer populated region, each feature is interrupted by exactly $\nu_{tot}$ vertical features, as schematically illustrated in Fig.1(g). This is consistent with discrete filling of the LLs in each layer: with increasing $D$, $\nu_T$ increases step-wise from $0$ to $\nu_{tot}$, and $\nu_B$ decreases from $\nu_{tot}$ to $0$. 
Interlayer charge transfer is allowed only when the displacement field induces crossings of LLs in the two layers.
At the crossing points, the bilayer system manifests increased polarizability at both integer and non-integer fillings, giving rise to the vertical features in Fig.1(e) (see SI.4 for more details).

\begin{figure*}[htb]
\begin{center}
\includegraphics{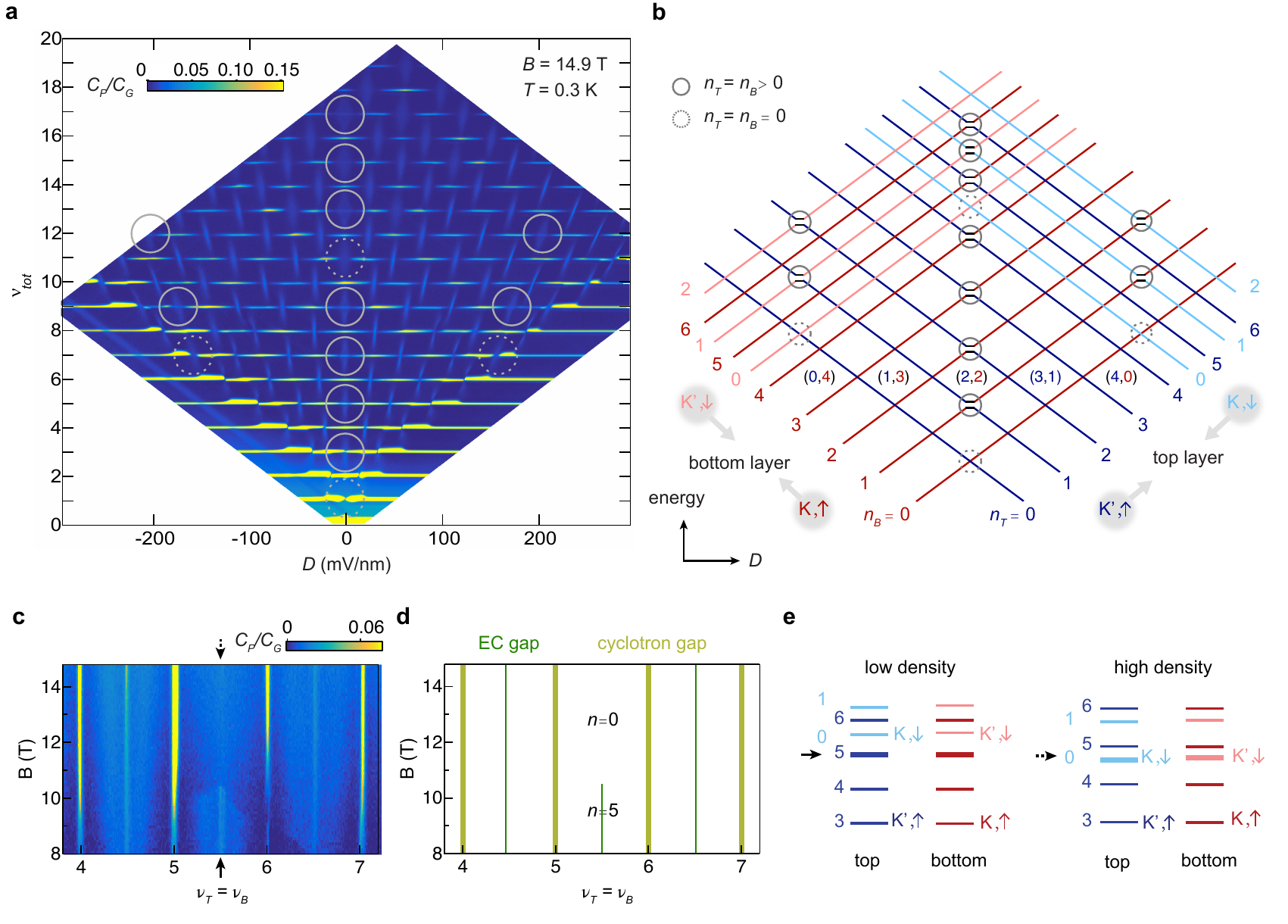}
\vspace{-0.15 in}
\end{center}
\caption{\small{
{\bf{Conditions for exciton condensate formation.}}
(a) Penetration capacitance versus total filling factor $\nu_{tot}$ and displacement field $D$ at $B = 14.9$ T. 
Solid (dotted) circles mark the crossings where the two LLs have the same orbital number $n = n_T = n_B > 0$ ($n = n_T = n_B = 0$).
(b) Schematic LLs diagram. As $D$ is varied, LLs in the bottom layer increases in energy and those of the top layer increases, giving rise to LL crossings.
Each layer hosts LLs from two spin branches, with a large spin-splitting energy several times of the cyclotron gap.
LLs from 4 different flavors are marked by different colors and the orbital number is also marked.
(c) Penetration capacitance at the layer balanced condition $\nu_T = \nu_B$, as a function of the filling factor and magnetic field.
(d) Schematic illustration according to (c), with yellow lines standing for the IQH gap, and the green lines standing for the EC gap. Orbital numbers are marked for different regions for $\nu_T = \nu_B = 5.5$.
(e)Cartoon illustrating the LL structure at low and high densities respectively. The black arrow points to the Fermi level at $\nu = 5.5$, where active LLs have orbital number $n = 5$ and $n = 0$, respectively.
}
}
\label{fig3}
\vspace{-0.15 in}
\end{figure*}

Next, we examine the penetration capacitance at the LL crossings.
Fig.~2 shows representative examples of the three types of behaviour we observe. 
At the center of each $\nu_{tot} - D$ map, a light blue diamond area demarcates the region where both layers host fractional filling factors and the system has finite polarizability.
Fig.2(a) and (b) show two examples where the gap closes, leading to a disappearance of the sharp horizontal feature at integer $\nu_{tot}$. This behavior is seen for most LL crossings, and is consistent with the single-particle picture of LL crossings - the bilayer system is compressible since both layers are partially filled. 
The gap closing in (a) and (b) also confirms the suppression of interlayer tunneling, which would otherwise induce LL anti-crossings and preserve the gap.
Fig.2(c), on the contrary, demonstrates an example where the gap persists throughout the fractional filling region.
The two scenarios regarding the presence or absence of a gap at the level crossings are clearly distinguishable in linecuts taken through the center of the fractional-filling region (bottom panels): a sharp spike in $C_P$ is seen at integer $\nu_{tot}$ for (c), but is absent in (a) and (b). 
Having established that the two layers are tunnel-decoupled, we interpret the incompressible states while both layers are partially filled in (c) as interlayer correlated states, \textit{i.e.} exciton condensates.

In Fig.3(a) we plot $C_P$ as a function of $\nu_{tot}$ and $D$ over a broad range, and mark the location of EC by solid circles using the criteria established in Fig.2. 
Remarkably, EC appear for a large range of LLs - far beyond what has been previously observed - both where the layers are balanced (around $D = 0$) and where they are very imbalanced (at large $D$). Fig.3(b) shows the corresponding schematic LL diagram, which identifies the LL orbital and spin/valley indices (see SI.2). We note first that all of the EC correspond to states with matched orbital number $n$. The series of EC at $D = 0$ (with Fig.2(c) as an example) corresponds to the crossings of two LLs with the same orbital number and spin indices but different valley index; while those at finite $D$ correspond to matched orbital numbers and valley indices but different spins (see SI.5).

\begin{figure*}
\begin{center}
\includegraphics{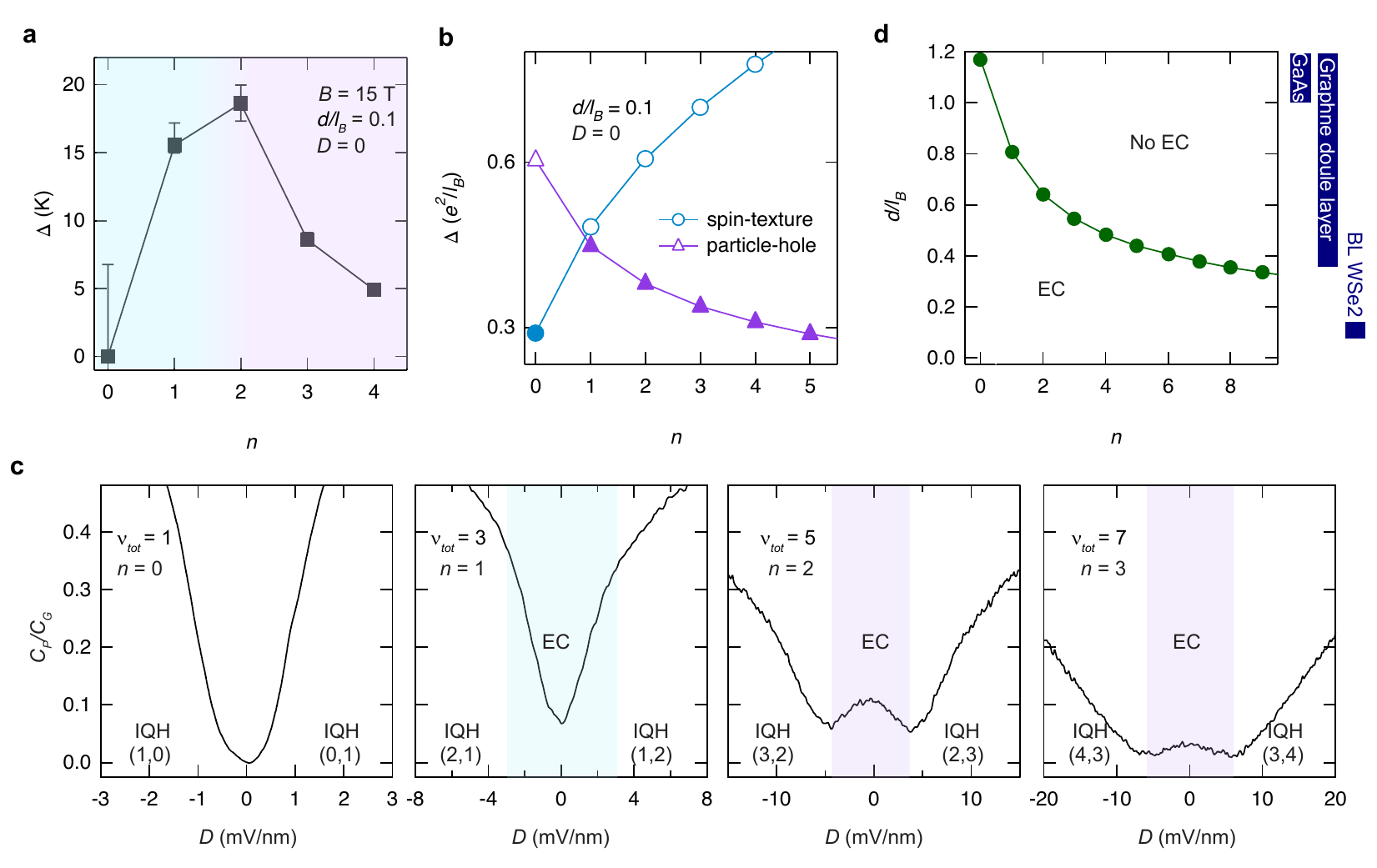}
\vspace{-0.15 in}
\end{center}
\caption{\small{
{\bf{Different types of charged excitations for different LLs.}} 
(a) Measured excitation gap when the layers are balanced ($D$ = 0 and LLs in both layer are half-filled), at $B$ = 14.9 T which corresponds to $d/l_B$ = 0.1. 
Error bar for $n =0$ reflects the uncertainty in distinguishing the EC gap at density balance from that at density imbalance or the IQH gap. 
Error bars for the rest represent the uncertainty in separating the stray capacitance or the polarizability contribution. 
Error bars are not shown when they are smaller than the symbol.
(b) Theoretical estimate of charge gap values for layer balanced condition, for two types of excitations, at $d/l_B = 0.1$. The excitation energy of layer-pseudospin texture increases with $n$, while those for particle-hole excitations decreases with $n$.
The filled markers stand for the lowest-energy excitations, which changes from one type to the other as $n$ varies.
(c) $C_P/C_G$ as a function of $D$, at constant total filling factor $\nu_{tot}$ = 1, 3, 5, 7, as marked, while the topmost LLs has orbital numbers $n$ = 0, 1, 2, 3, respectively.
A small background due to the polarizability contribution has been subtracted according to $C_P$ at adjacent non-integer values.
The shades highlight the $D$-range where the both layers are partially filled and charge is being transferred interlayer, while in the non-shaded regions both layers are at integer fillings as marked and the system exhibits an IQH gap.
(d) Numerically calculated upper limit $d/l_B$ for EC formation. Experimentally accessible ranges of $d/l_B$ for different systems are marked on the right.  
}}
\label{fig4}
\vspace{-0.15 in}
\end{figure*}

Our observations indicate that EC formation requires the orbital wavefunctions to match, but is independent of the spin/valley degree of freedom.
Likewise, our Hartree-Fock and numerical calculations suggest that formation of EC, characterized by a good overlap with the Halperin (111) wavefunction and a gapless neutral mode, only appears when the LLs in the two layers have matched orbital numbers (see SI.10 and SI.11).
The consistency with theory further confirms the interpretation of the observed incompressible states as EC. In addition, our observation is consistent with previous experiments in graphene double-layers \cite{li:2017a,li:2019} which demonstrated that spin and valley degrees of freedom are largely irrelevant in the formation of EC in the lowest LL. Intriguingly, the specific orbital number also plays a critical role in EC formation: EC formation is observed for $n = 1-6$, but is suppressed for $n = 0$, as shown by the dashed circles in Fig.3 (a) and (b) (see SI.5 and SI.6). This is precisely opposite to previous studies in double-layer systems with larger interlayer spacing \cite{eisenstein:2014,li:2017a,liu:2017}, where the EC have only been observed for $n = 0$.

The critical role of the orbital number is further corroborated by the evolution of specific EC with magnetic field. 
This is displayed in Fig.3(c), which plots $C_P$ versus the magnetic field and filling factor at the balanced condition $D = 0$. For $B < 10.5$ T, EC are clearly observed for $\nu_T = \nu_B  = 4.5$, $5.5$, and $6.5$, but the state at $\nu_{B,T}=5.5$ disappears abruptly above $10.5$ T. This disappearance of EC coincides with a change in the orbital number from $n = 5$ to $n = 0$, as illustrated in Fig.3(d).
The orbital number change is due to the rearrangement of the LLs with different spin indices, as illustrated in Fig.3(e). 
For a fixed filling factor, the carrier density scales linearly with the magnetic field, and at low densities, the $g$-factor which decides the Zeeman splitting between two spin branches are strongly enhanced \cite{shi:2020}.
Therefore, as $B$ increases and the carrier density is raised above a threshold value, the active LLs at a fixed filling factor can switch from one spin to another and abruptly change orbital number to $n=0$, leading to the disappearance of EC (see SI.7).

To better understand the dependence of EC robustness on orbital number, we measure the excitation gap $\Delta$ by integrating $\partial \mu/\partial n$ over the gap at integer filling factors (see SI.8 for details).
Fig.4(a) plots $\Delta$ \textit{vs.} the orbital number $n$ at $D = 0$, for $B = 14.9$ T: it exhibits a non-monotonic behavior with the maximum at $n = 2$.
Tuning the orbital number modifies the single-particle electron wavefunction:
as $n$ increases, the wavefunctions spread out and have more nodes. In single-layer systems, at fractional fillings, this evolution often leads to different correlated ground states for different values of $n$. 
Here, although the ground state remains EC, the non-monotonic dependence of $\Delta$ on $n$ stems from the change of the nature of low-energy charged excitations. The character of these excitations is more conveniently understood in the pseudospin picture.
In this picture, electron states in the top (bottom) layer are viewed as pseudospin up (down), while the phase-coherent EC manifest macroscopically aligned pseudospin pointing in-plane \cite{moon:1995}.
On top of such a pseudospin ferromagnet ground state, one type of charged excitation is associated with a spatially extended pseudospin texture, known as meron-antimeron pair in the case of bilayer quantum Hall systems~\cite{moon:1995} (similar to skyrmions~\cite{sondhi:1993} in quantum Hall ferromagnets in a single-layer).
As $n$ increases, the energy of such excitations increases~\cite{wu:1995}.
On the other hand, a conventional type of excitation, which represents a localised particle or hole, has an energy that decreases with $n$. As a result, a change in the nature of the lowest energy excitations is expected as $n$ is varied.
Fig.4(b) shows the theoretical estimate of the gap for the two types of excitations in a simple bilayer model without screening effects. 
The transition between two types of excitations occurs between $n = 0$ and $n = 1$, giving a non-monotonic dependence. This is qualitatively consistent with our data, except that in the model the gap maximum appears at $n=1$, whereas the maximum appears at $n=2$ in our measurement, suggesting that the pseudospin textured excitations also dominate at $n = 1$.
The slight discrepancy with experimental data may be due to screening or disorder effects, which may lower the energy of pseudospin-texture excitations and shift the transition to higher $n$.

The different characteristics of the two kinds of excitations are also suggested by the density imbalance dependence, as shown in Fig.4(c).
Here, we plot the penetration capacitance as a function of $D$, at a constant total filling factor $\nu_{tot}$ = 1, 3, 5, 7, which are the four bottom circles in Fig.3(a) and (b), and correspond to LL orbital number $n$ = 0, 1, 2, 3 respectively.
The color shades in the right three panels illustrate the $D$-range where the individual layers host fractional filling factors, as suggested by the polarizability contribution at adjacent non-integer fillings.
A larger $C_P$ suggests a smaller electronic compressibility and a larger gap for charged excitations.
Our data suggests that, for $n = 1$, the EC gap increases with $D$, and transitions smoothly into the IQH gap.
In contrast, for $n$ = 2 and 3, the EC gap decreases with $D$, and has a minimum before transitioning into the IQH state.

The distinct behavior of the two types of excitations is supported by our numerical calculations (see SI.12), which find that the meron-antimeron pairs have a sharp increase in gap with layer imbalance, while the particle-hole excitations shows a flat response.
While the pseudospin points in-plane for a density-balanced bilayer system, imbalancing the bilayer is equivalent to tilting the pseudospin out-of-plane, therefore the in-plane component of the spin-stiffness decreases \cite{moon:1995}.
For meron-antimeron pairs, the excitation energy has the main contribution from the meron and anti-meron self-energy, the sum of which increases with density imbalance.
On the other hand, the particle-hole excitation energy is just the exchange energy, which is independent of the charge density imbalance.
The decrease of the excitation gap with imbalance for $n = 2$ and 3 is not captured by our calculations and remains to be understood.

Finally, we remark that, in contrast to our bilayer \wse2, in other quantum Hall double layer systems the EC has only been observed within the LL of orbital number $n = 0$.
We find that, although numerical calculations suggests that EC could form in LLs of high orbital numbers, a smaller interlayer spacing is required for higher orbital numbers. 
As shown in Fig.4(d), the calculated critical interlayer spacing in units of the magnetic length, $d_c/l_B$, decreases sharply as a function of the orbital number $n$. The critical spacing $d_c$ is estimated as the distance where the Goldstone mode goes soft at finite momentum, signaling the EC becoming unstable (see SI.11).
On the right axis of Fig.4(d), we mark the experimentally accessible ranges of $d/l_B$ for three different systems -- GaAs double quantum wells, graphene double layers with BN barrier, and bilayer \wse2.
Our calculation suggests that the absence of EC in higher LLs in the first two systems is likely due to the large thickness of the intentional interlayer physical barrier (graphene also has the complication as the LL wavefunctions are mixture of simple harmonic oscillator wavefunctions).
In contrast, the interlayer spacing in bilayer \wse2 is set by the lattice constant in the vertical direction, which is only about 0.7 nm, and thus EC are expected to occur in higher LLs.

To conclude, our study demonstrates bilayer \wse2 as a unique platform to access EC in the strong coupling limit, by exploiting the intrinsic spin-orbital coupling and stacking order in the host material. This designing principle could be further used in creating other quantum phases from layered van der Waals materials, including the EC in zero field. Moreover, our study demonstrates EC composed from tunable single-particle orbital wavefunctions, and how such tunability could vary the properties of EC and the excitations on top of the ground state.


We thank William Coniglio and Bobby Pullum for help with experiments at the National High Magnetic Field Lab.
This research is primarily supported by US Department of Energy (DE-SC0016703).
Synthesis of \wse2 (D.R.,B.K.,K.B.) was supported by the Columbia University Materials Science and Engineering Research Center (MRSEC), through NSF grants DMR-1420634 and DMR-2011738. 
A portion of this work was performed at the National High Magnetic Field Laboratory, which is supported by National Science Foundation Cooperative Agreement No. DMR-1157490 and the State of Florida. D.A. acknowledges support by the Swiss National Science Foundation and by the European Research Council (grant agreement No. 864597). Z.P. acknowledges support by the Leverhulme Trust Research Leadership Award RL-2019-015. K.W. and T.T. acknowledge support from the Elemental Strategy Initiative conducted by the MEXT, Japan (Grant Number JPMXP0112101001) and JSPS KAKENHI (Grant Numbers JP19H05790 and JP20H00354).

\section{Method}

The heterostructure is assembled using the van der Waals dry transfer technique, and pre-patterned Pt electrodes are used for electrical contacts to bilayer \wse2.
Hexagonal boron nitride are used as the dielectric, and graphite or metal as top and bottom gates for bilayer \wse2.
The top gate was lithographically shaped so that its overlap with bottom gate covers only bilayer \wse2, and the overlap defines the device area.
In order to achieve good electrical contact, we use an additional contact gate on top to heavily dope the contact area, which is isolated from the top gate by Al$_2$O$_3$ dielectric.
Data from a different device are shown in SI.9.
Penetration capacitance was measured with an FHX35X high electron mobility transistor serving as a low-temperature amplifier, in a similar setup as in \rref{shi:2020}.
Measurements were performed at $T = 0.3$ K unless otherwise specified.


\newpage

\newpage
\clearpage

\pagebreak

\begin{center}
\textbf{\large Supplementary Information for ``Bilayer WSe$_2$ as a natural platform for interlayer exciton condensates in the strong coupling limit"}\\

\end{center}


\section{S1. Discussion on interlayer tunneling}

Here we discuss the suppression of interlayer tunneling in the context of conventional double quantum wells.
In conventional GaAs double quantum wells, the interlayer tunneling is usually characterized by $\dsas$. The two single-particle eigenstates can be written as,

\begin{align}
\psi_{+} &= \sin \alpha\; \psi_{t} + \cos \alpha\; \psi_{b}\,,\\
\psi_{-} &= \cos \alpha\; \psi_{t} - \sin \alpha\; \psi_{b}\,.
\end{align}

Here, $\phi_{t}$ and $\phi_{b}$ are the wavefunctions localized in the top and bottom quantum wells, respectively;
\be
\cos 2\alpha = eV/\sqrt{(eV)^2+\dsas^2}
\ee
characterize the degree of wavefunction localization, where $V$ is the interlayer electrical potential difference.

For any finite tunneling gap $\dsas$, $\cos 2 \alpha = 0$ at zero bias and the single-particle wavefunctions are just the bounding and anti-bounding states that is an superposition of wavefunctions in both layers with equal weight ($\cos \alpha = \sin \alpha$). 
The single-particle ground state has a gap $\dsas$ to the excited state.

In contrast, the bilayer \wse2 manifest a large degree of wavefunction localization in individual layers even at zero bias, and this is a result of the strong spin-orbit coupling. 
It was shown that the spin doublet single-particle eigenstates in the valence band can be written as \cite{gong:2013},

\begin{align}
|\text{K},\uparrow\rangle &= (\sin\beta\; \psi_t^{\text{K}} + \cos\beta\; \psi_b^{\text{K}}) \otimes |\uparrow\rangle\,,\\
|\text{K},\downarrow\rangle &= (\cos\beta\; \psi_t^{\text{K}} + \sin\beta\; \psi_b^{\text{K}}) \otimes |\downarrow\rangle\,,\\
|\text{K}',\downarrow\rangle &= (\sin\beta\; \psi_t^{\text{K}'} + \cos\beta\; \psi_b^{\text{K}'}) \otimes |\downarrow\rangle\,,\\
|\text{K}',\uparrow\rangle &= (\cos\beta\; \psi_t^{\text{K}'} + \sin\beta\; \psi_b^{\text{K}'}) \otimes |\uparrow\rangle\,,
\label{wse2tn}
\end{align}
where 
\begin{align}
    \psi_t &= \frac{1}{\sqrt{2}}(|d_{x^2-y^2}^t\rangle-i\tau|d_{xy}^t\rangle)\,\\
    \psi_b &= \frac{1}{\sqrt{2}}(|d_{x^2-y^2}^b\rangle+i\tau|d_{xy}^b\rangle)
\end{align}
is the wavefunction localized in top and bottom layers and $\tau$ is the valley index (1 for K and -1 for K'),
\be
\cos 2 \beta = \lambda/\sqrt{\lambda^2+t^2}\,,
\ee
$\lambda$ is the spin-orbit coupling strength and $t$ is the interlayer hopping amplitude. 

For holes in bilayer \wse2, $\cos 2 \beta \approx 0.96$ according to ab initio calculations \cite{gong:2013}. 
Therefore, the single-particle eigenstate wavefunctions are always highly localized in individual layers.
The spin-orbital coupling strength $\lambda$ is effectively an intrinsic bias and only the low energy branches are accessible in our experiment.
All four flavors described in equations above are degenerate in energy in the absence of external fields, and no single-particle gap is expected. 
Such energetically degenerate yet spatially separated single-particle wavefunctions set the basis for exciton condensate physics.

\begin{figure*}
\begin{center}
\includegraphics[width=\linewidth]{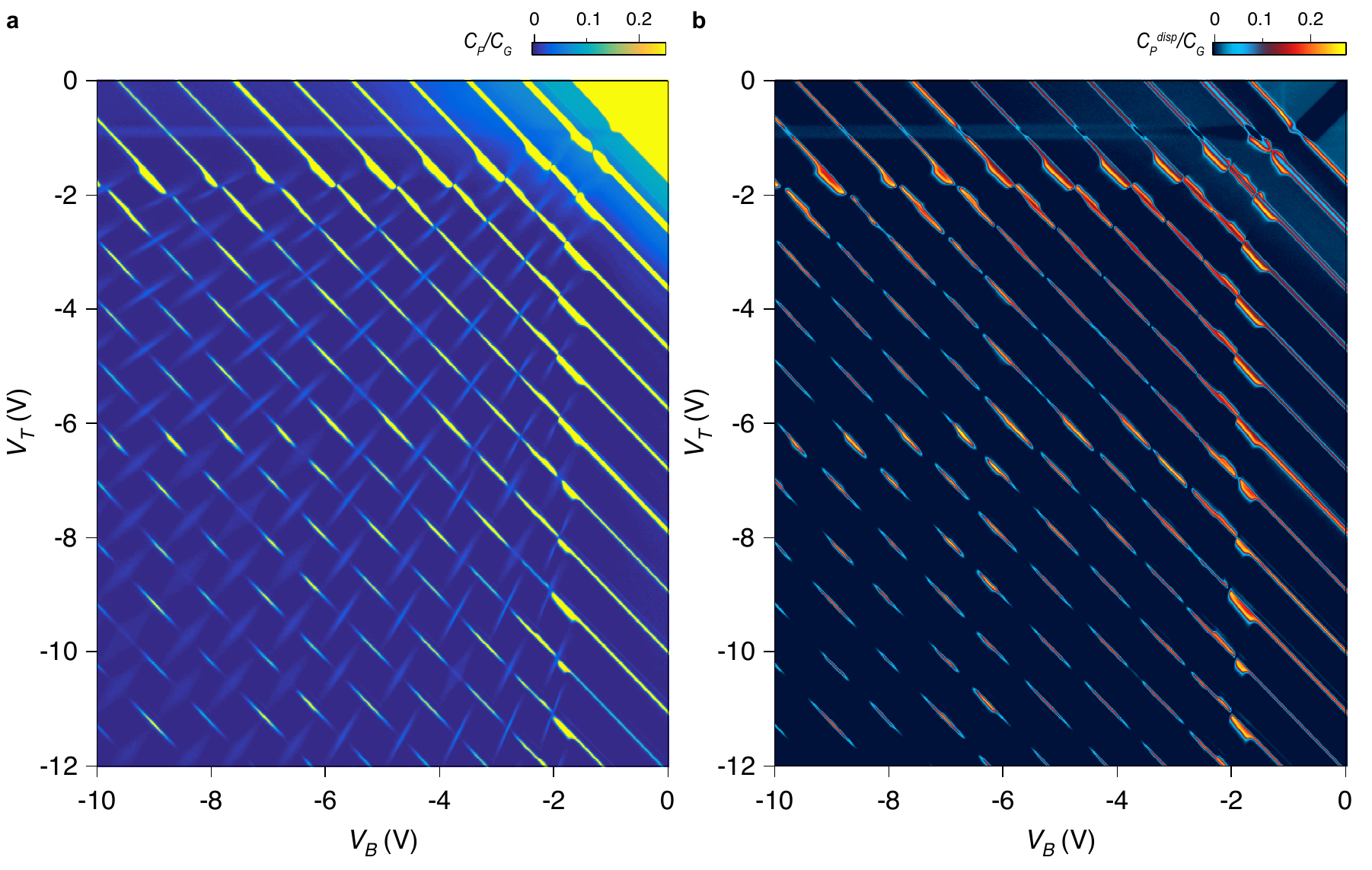}
\vspace{-0.15 in}
\end{center}
\caption{
\small{
(a) Capacitive and (b) dissipative contributions in the penetration capacitance measurements as a function of top and bottom gate voltages $V_T$ and $V_B$.
}
}
\label{polar}
\vspace{-0.15 in}
\end{figure*}

\section{S2. Landau level structure and wavefunctions in bilayer \wse2}

The above discussion suggests a high level of layer localization which persists to the LL regime.
We therefore treat the LL structures in the two layers independently.
In a monolayer \wse2, the LL structure has two isospin flavors, separated by a large spin-splitting energy.
The spin susceptibility, characterized by $E_Z/E_C$, where $E_Z$ is the spin-splitting energy and $E_C$ the cyclotron energy, increases as carrier density decreases and could be as large as 9 \cite{gustafsson:2018,shi:2020}.
In bilayer \wse2, the same physics applies, but due to a different dielectric environment with larger dielectric constant, the spin susceptibility is slightly smaller.
Further, the susceptibility in one layer mainly depends on the carrier density of this layer, and slightly on the density of the other layer.
In our experiment, the spin of a certain LL is decided from similar analysis as in \cite{shi:2020}, and assisted with transport measurements where different spin flavors exhibits very different conductivity \cite{lin:2019}.
At 15 T, five spin-up LLs are populated before the first spin-down LL is populated in each layer.

It is well known that for the valence band of \wse2, the LL with orbital index $N = 0$ only exists in one valley; in the other valley, the lowest LL has orbital index $N = 1$ \cite{rose:2013}.
However, it is worth noting that the LL wavefunctions is not related to the orbital index in the simple way as in non-Dirac systems, but a mixture of the harmonic oscillator wavefunctions. 
Indeed, in the valley with LL orbital index $N = 0$, the wavefunction is pure $|0\rangle$, while the next LL with $N = 1$ is mainly $|1\rangle$ with a few percent of mixing of $|0\rangle$, where $|n\rangle$ stands for the harmonic oscillator wavefunction of index $n$.
On the other hand, in the valley without a $N = 0$ LL, its lowest LL with $N = 1$ actually has the wavefunction dominated by $|0\rangle$ and only with a few percent of mixing from $|1\rangle$.
Therefore, to the first order, the lowest LLs in the two valleys have similar wavefunctions despite of different LL orbital index. 
This is important to establish interlayer EC in the case of mismatched spins.
In the main text, the orbital number $n$ stands for the index associated with the dominant component $|n\rangle$ of the single-particle wavefunction.

\begin{figure*}
\begin{center}
\includegraphics{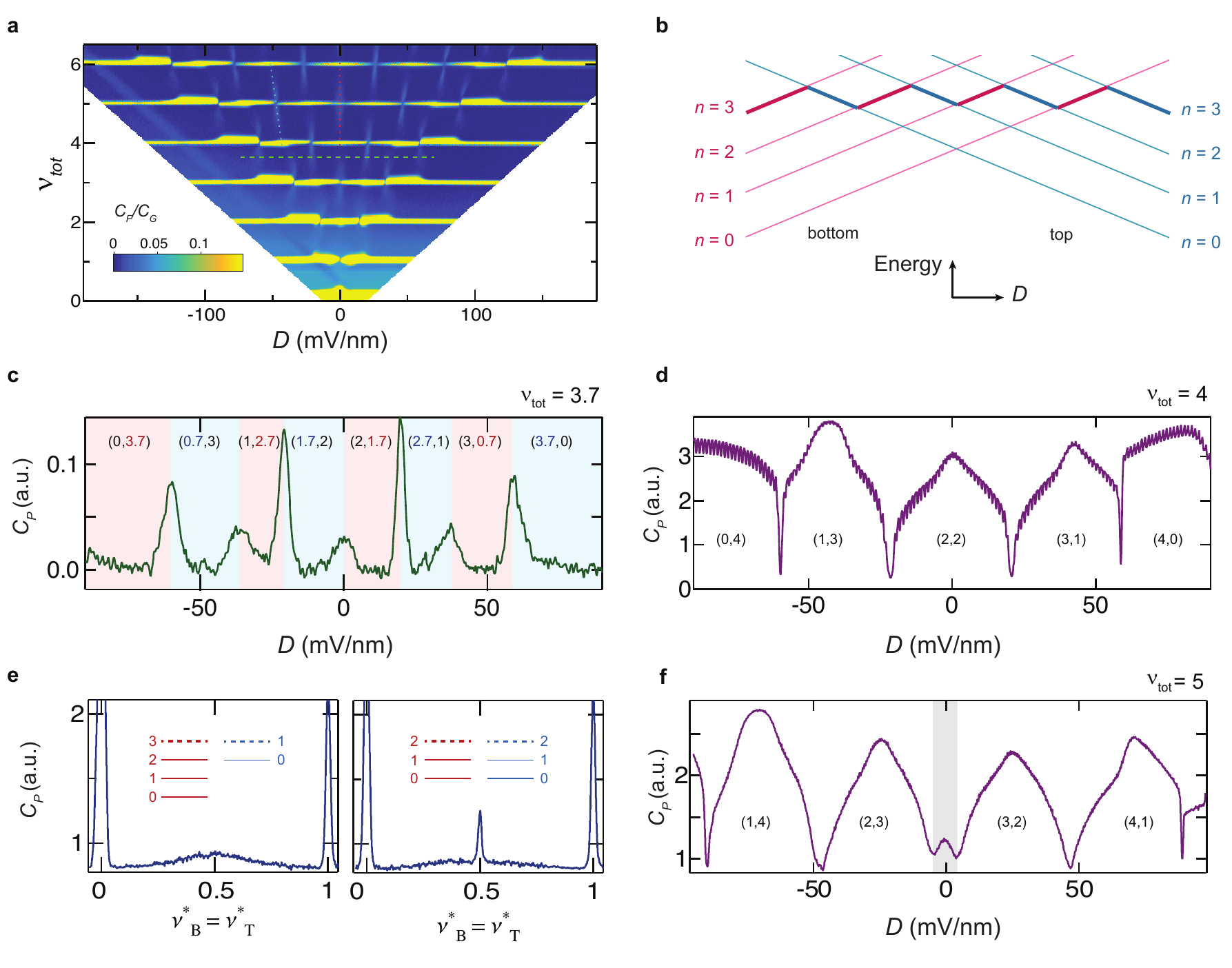}
\vspace{-0.15 in}
\end{center}
\caption{\small{
(a) Penetration capacitance versus total filling factor $\nu_{tot}$ and displacement field $D$ at $B = 14.9$ T, focusing on the low $\nu_{tot}$ regime.
(b) Schematic LLs diagram. As $D$ is varied, LLs in the bottom layer increases in energy and those of the top layer decreases, giving rise to LL crossings.
(c) $C_P$ \textit{vs.} $D$ at $\nu_{tot} = 3.7$ along the green line in (a). Blue and red shades mark the region where the Fermi level is in the top and bottom layer, respectively, with the filling factor marked as $(\nu_T,\nu_B)$.
(d) $C_P$ \textit{vs.} $D$ at $\nu_{tot} = 4$.
The dips mark where the gap closes as a result of LL crossings.
(e) $C_P$ \textit{vs.} filling factor along $\nu^*_B = \nu^*_T$. The left and right panel are along the blue and red lines in (a), respectively. In the left panel the valence LLs are has orbital numbers $n_B = 3$ and $n_T = 1$, while in the right panel $n_B = n_T = 2$. In the latter case, a peak is observed at when $\nu^*_T + \nu^*_B = 1$, signaling gap opening.
(f) $C_P$ \textit{vs.} $D$ along $\nu_{tot} = 5$.The grey shaded region is where both layers host partial filling factors. 
}}
\label{linecuts}
\vspace{-0.15 in}
\end{figure*}


\section{S3. Negative compressibility}

The strong correlation effects in TMDs result in a negative compressibility which is stronger at lower densities \cite{pisoni:2019b}.
This effect manifests in several aspects of our data.
In Fig.1 (d) and (f) of the main text we outline the polarization of the bilayer at zero field.
Here the boundary of the fully polarized region is curved.
In a simple electrostatic model, 
\be
E = (D - e\Delta n /2)/\epsilon\,,
\ee
where $\epsilon$ is the dielectric constant and $E$ is the unscreened interlayer electrical field.
Without interaction effects considered, $E$ would not have any dependence on the total density, and the displacement field required to fully polarize the bilayer should increase linearly with the total density.
Due to the negative compressibility, $E$ is of opposite sign to $D$, and its amplitude increases with decreasing $n$ - the bilayer electrons tend to build a larger density imbalance to``overscreen" the displacement field.
Therefore, the displacement field required to fully polarize the bilayer increases superlinearly with the total density, giving rise to the curve shaped boundary in Fig.1 of main text.

Similarly, the negative compressibility effect also manifest in the data under a perpendicular magnetic field.
Under a magnetic field, a single-particle picture would predict that the LL crossings happen at the same $D$ for a certain filling factor difference $\Delta\nu = \nu_B - \nu_T$, as shown in Fig.3(b) in the main text.
In contrast, as shown in Fig.3(a) in the main text, for a fixed $\Delta\nu$, the LL crossings happen at a smaller $D$ for smaller $n$, and the polarizability segments at partial fillings also assume a slope.

\section{S4. Distinguishing polarizability and incompressibility contribution}

The penetration capacitance is a sum of both ``incompressibility" and ``polarizability" contributions.
First, we confirm that the polarizability contribution is only in the capacitance channel, and is not dissipation related.
In \rfig{polar} we plot the capacitive and dissipative signal in the measured penetration capacitance in (a) and (b), respectively.
While most integer quantum Hall gaps show up in both, the features cutting the integer gap only appear in the capacitive channel, which are consistent with the polarizability contribution.

We further distinguish the two contributions according to their behavior at different filling factors.
\rfig{linecuts} (b) shows the schematic LL diagram corresponding to the data in \rfig{linecuts} (a).
In \rfig{linecuts}(c) we plot the linecut of $C_P$ along constant non-integer filling factor $\nutot = 3.7$ [horizontal line in \rfig{linecuts}(a)].
Here, the Fermi level changes back and forth between two sets of LLs, as marked by the thick lines in \rfig{linecuts}(b).
The bilayer system is always compressible and the peaks arises from the polarizability contributions .
Indeed, whenever the Fermi level changes between the LLs localized in different layers, interlayer charge transfer happens, and gives rise to a peak in $C_P$.
We observe a total of seven peaks in the whole $D$ range in \rfig{linecuts}(c), separating red and blue regions, which correspond to the cases that the Fermi level is in the bottom (red lines) and top (blue lines) layers, respectively.
Further, a small peak corresponds to charge transfer amount of 0.3 of the Landau level degeneracy, while a large peak corresponds to 0.7, marked by the numbers in \rfig{linecuts}(c).

In \rfig{linecuts}(d) we plot the linecut of $C_P$ along constant integer filling factor $\nutot = 4$.
We observe that $C_P$ show four dips in the whole $D$ range, which corresponds to four LL crossings, separating regions of different filling factors marked in \rfig{linecuts}(d).
At the level crossings, in principle there is ``polarizability" contribution as well, but it is much smaller, as evidenced from different vertical scales for \rfig{linecuts}(c) and (d). 

To distinguish the two contributions at LL crossings at integer filling factors, we make a linecut along the polarizability segments, cutting through integer filling lines, as shown in \rfig{linecuts}(e).
Such a linecut represents the condition $\nu_T^* = \nu_B^*$, where $\nu_T^*$ ($\nu_B^*$) is the fractional part of the filling factor in the top (bottom) layer.
While the polarizability contribution continuously changes from non-integer $\nutot$ to integer $\nutot$, an incompressible contribution would only be possible at integer $\nutot$, and shows up as a sharp peak at $\nu_T^* =\nu_B^* = 0.5$.
The two panels in \rfig{linecuts}(e) demonstrate the two situations, respectively. 
In the right panel, an EC gap exist as orbital numbers match in the two layers; in the left panel, an EC gap does not exist as the orbitals are mismatched.

In the presence of EC, the linecut along constant $\nu_{tot}$ is shown in \rfig{linecuts}(f).
Here, while $C_P$ at other LL crossings demonstrate simply a dip at the closing of IQH gap, its behavior in the EC (grey shaded) region is different, exhibiting an ``reentrant" increase of $C_P$, which corresponds to the emergence of a correlation gap.

\section{S5. Exciton condensate at density imbalance}

\begin{figure*}
\begin{center}
\includegraphics{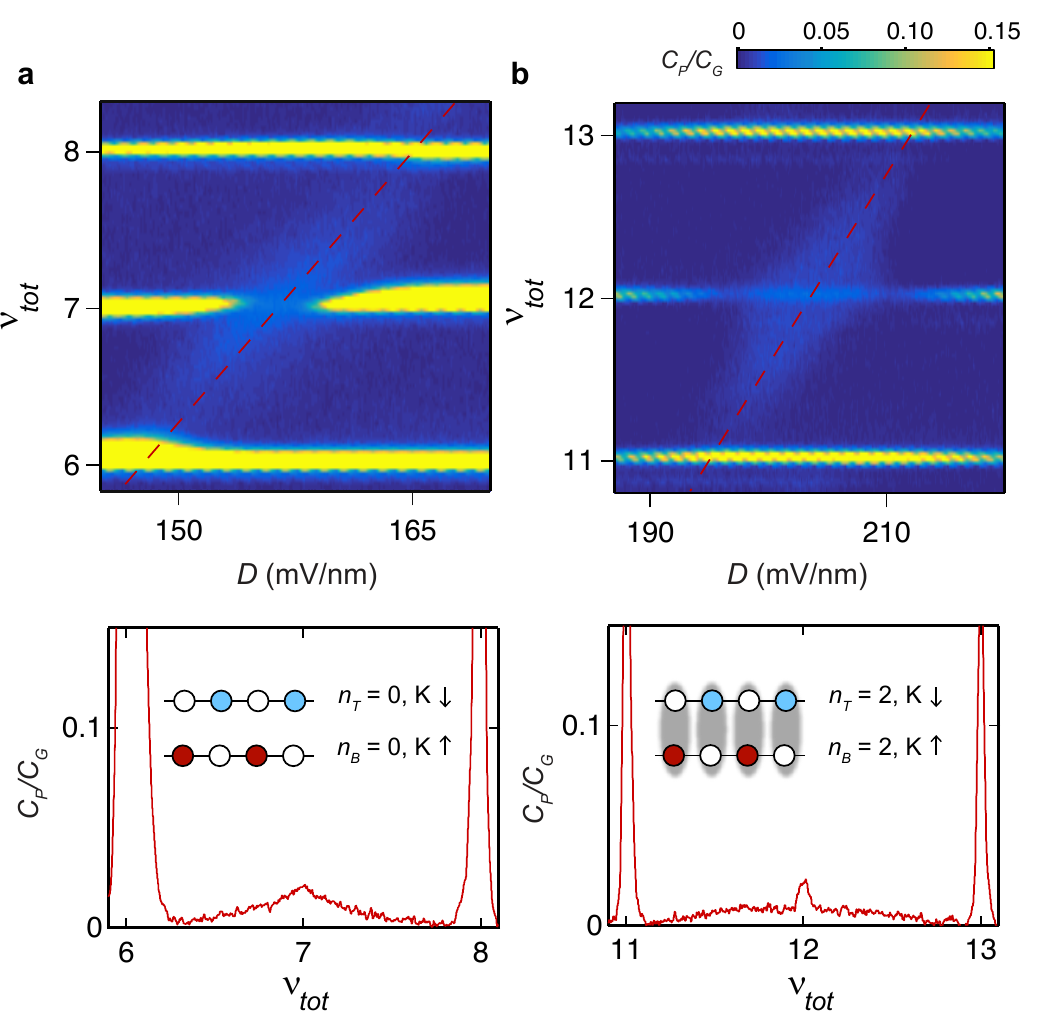}
\vspace{-0.15 in}
\end{center}
\caption{\small{
Penetration capacitance versus total filling factor $\nu_{tot}$ and displacement field $D$ at $B = 14.9$ T around LL crossings of opposite spins.
The bottom panel are linecut along the red dashed line in the top panel, which tracks equal topmost LL population in the two layers.
The inset of the bottom panel illustrates the orbital, spin and valley indices.
}}
\label{maphrdp}
\vspace{-0.15 in}
\end{figure*}


In \rfig{maphrdp} we plot $C_P/C_G$ for LL crossings with the same orbital number but different spin.
For $n = 0$, as shown in \rfig{maphrdp} (a), no incompressible state is observed when both layers host partially filled LLs.
In contrast, for $n = 2$, as shown in \rfig{maphrdp} (b), incompressible state persists when both layers are partially filled, consistent with EC formation.
The suppression of EC at $n = 0$ is consistent with what we observe for the LL crossings between the same spins, at $D = 0$, demonstrated in Fig.3 of the main text.

\section{S6. Data for $n = 0$ at 30 T}

\begin{figure}[b]
\begin{center}
\includegraphics[width = \linewidth]{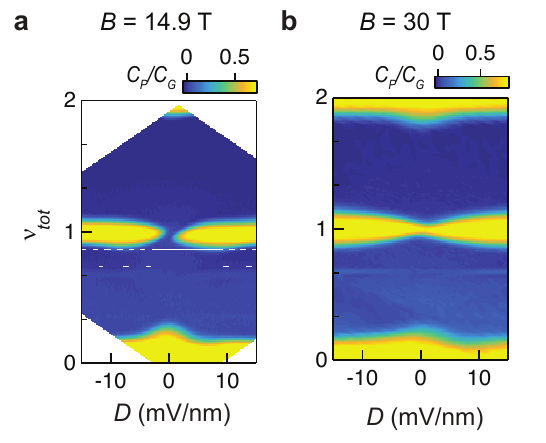}
\vspace{-0.15 in}
\end{center}
\caption{\small{
Penetration capacitance as a function versus $\nu_{tot}$ and $D$ at (a) $B$ = 14.9 T and (b) $B$ = 30 T. 
}}
\label{n0map}
\vspace{-0.15 in}
\end{figure}


In \rfig{n0map} (a) and (b) we show the capacitance as a function of top and bottom gate voltages at 14.9 T and 30 T, focusing at $\nu_{tot} = 1$.
The charge gap at $\nu_{tot} = 1$ disappears at the layer balanced point $D = 0$ at 14.9 T, while it recovers at 30 T.
The difference could be due to a larger interaction energy scale at 30 T, which enhances the EC gap that overcomes disorder effects.

\section{S7. Magnetic field dependence at density balance}

\begin{figure}
\begin{center}
\includegraphics[width=\linewidth]{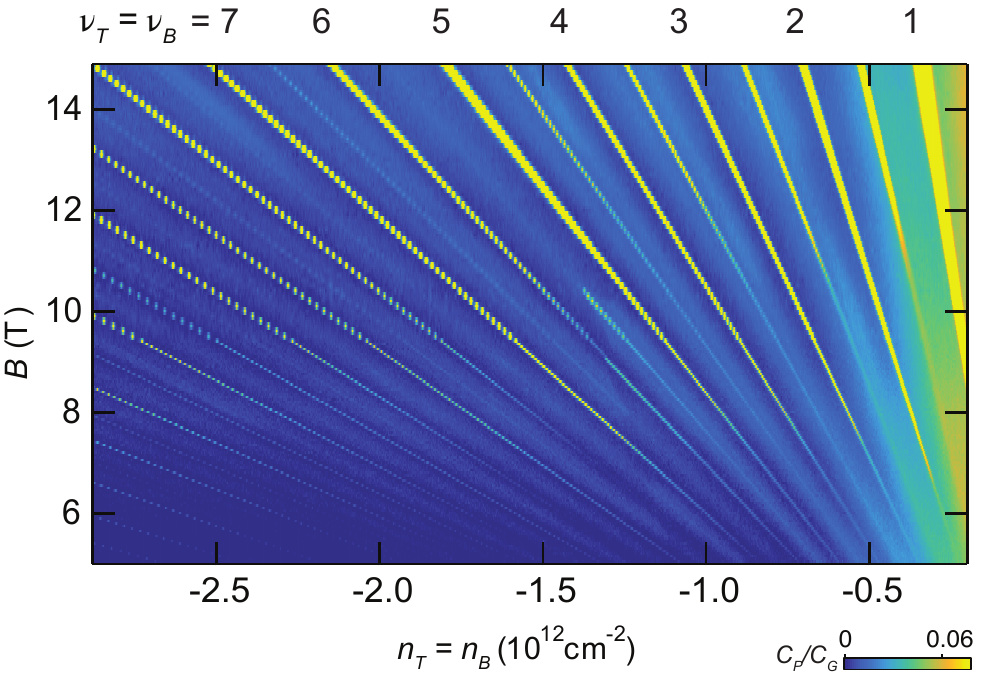}
\vspace{-0.15 in}
\end{center}
\caption{\small{
Penetration capacitance at the balanced density condition $n_T = n_B = n$, plotted versus $n$ and $B$.
Filling factor for individual layers are marked on the top.
}}
\label{fanchange}
\vspace{-0.15 in}
\end{figure}


In \rfig{fanchange}(a) we plot the same data as in Fig.3(c) in the main text, which shows $C_P$ at density balance as a function of the magnetic field, yet here versus the density and in a larger filling factor range.
The transition at $\nu_T = \nu_B = 5.5$ is clearly distinct from the behavior at other half-integer filling factors.
This transition originates from a density dependent $g$-factor, which varies the ratio of spin-splitting energy $E_z$ and the cyclotron energy $E_c$ \cite{gustafsson:2018,shi:2020} and shuffles LLs of different orbital and spin index as density is varied.

\section{S8. Extraction of charge gap}

For fixed excitation amplitude and frequency, the measured penetration signal $V_P$, to a good approximation, can be written as \cite{dultz:2000,goodall:1985},
\begin{equation}
    V_P^{complex} \propto C_P = C_G \lb 1-\frac{C_q}{C_t+C_b+C_q} \frac{\tanh(\alpha)}{\alpha} \rb \,,\\
      \label{maineq1}
\end{equation}
\begin{equation}
    \alpha = \sqrt{i \omega  \frac{C_q (C_t + C_b)}{C_t + C_b + C_q} R_s }\,,
    \label{maineq2}
\end{equation}
where the imaginary and real part, denoted $V_P$ and $V_P^{disp}$ stands for capacitive and dissipative contribution, respectively.
In the equation above, $C_t$ ($C_b$) is the geometry capacitance between \wse2 and the top (bottom) gate, $C_G = C_t C_b /(C_t + C_b)$ is the geometry capacitance between the two gates, $C_q = A e^2 (\partial \mu/\partial n)^{-1}$ is the quantum capacitance of the \wse2, with $A$ being the area of the device, and $R_s$ is the resistance of the \wse2 channel.

In the low frequency and low resistance limit ($\alpha \to 0)$, the real part - dissipative contribution is zero, and
the penetration capacitance $C_P$ can be written as,
\begin{equation}
    C_P = C_G \frac{C_t+C_b}{C_b + C_t + C_q}\,;
    \label{lowf}
\end{equation}
in the high resistance limit ($\alpha \to \infty)$ or when the \wse2 is gapped, the penetration capacitance is equal to the geometry capacitance between the two gates, 
\begin{equation}
    C_P = C_G\,,
    \label{gap}
\end{equation}
and the dissipative contribution is also zero.

\begin{figure}
\begin{center}
\includegraphics[width=\linewidth]{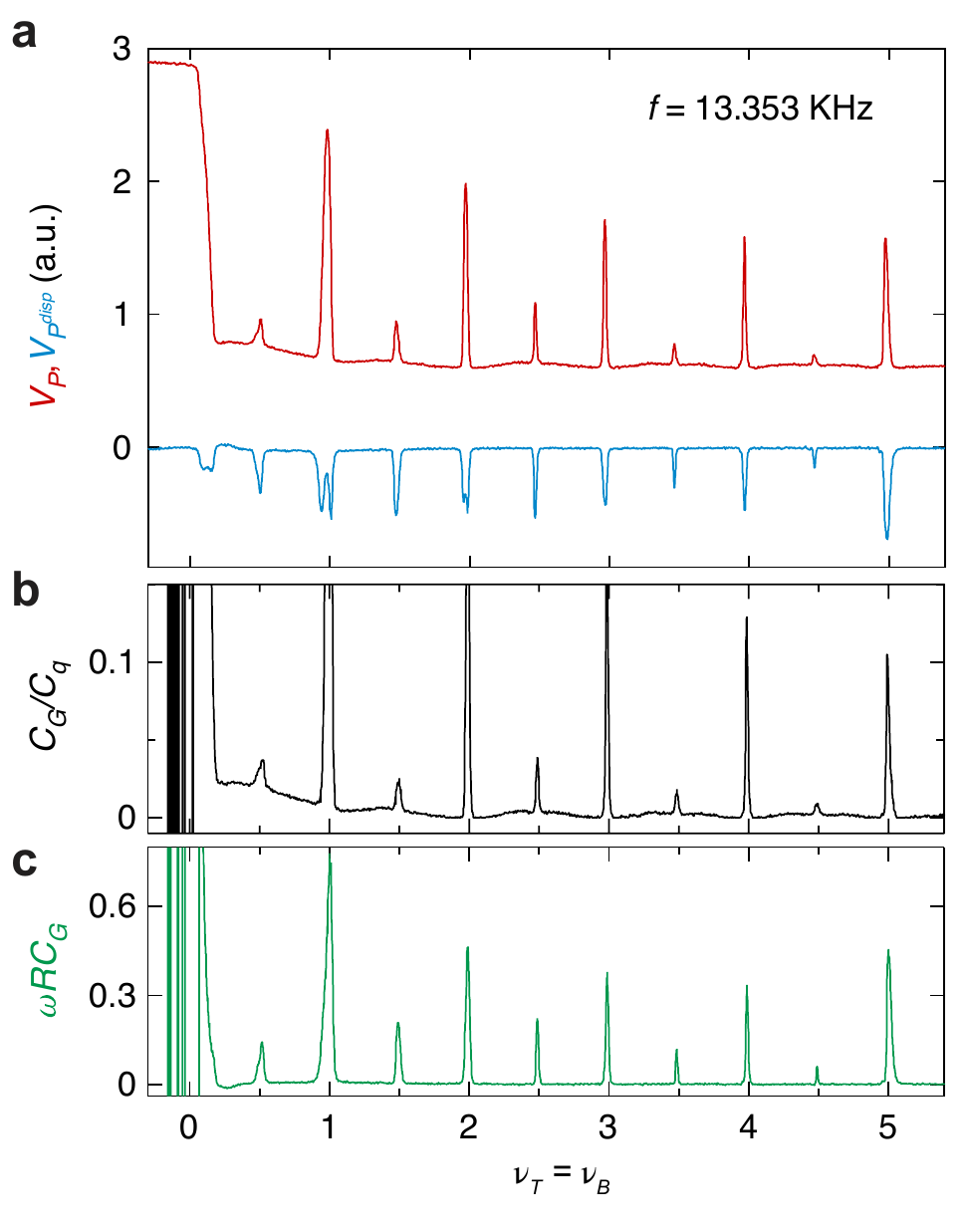}
\vspace{-0.15 in}
\end{center}
\caption{\small{
(a) Imaginary ($V_P$) and real ($V_P^{disp}$) part of the penetration signal measured at $B = 14.9$ T, as a function of the filling factor at the density balanced condition. For $\nu_B = \nu_T \approx 0.5$, the gap depends extremely sensitive on the total filling factor then filling factor imbalance, as shown in \rfig{n0map} (a), so the $n = 0$ gap is overestimated in this measurement due to a small density imbalance. In Fig.4 in the main text, we obtained the gap for $n = 0$ when the density balance condition is more accurately satisfied. (b) $C_G/C_q$ and (c) $\omega R C_G$ extracted from data in (a) by solving for \req{maineq1} and \req{maineq2}. The contact resistance is large at low densities, and may give rise to additional contribution for $\nu_B = \nu_T < 2$, which is not accounted for in our analysis. The gaps for $n = 0$ and $n = 1$ may thus be overestimated, but it doesn't change the qualitative trend, i.e., the gap has a non-monotonic dependence on $n$.
}}
\label{RC}
\vspace{-0.15 in}
\end{figure}

\begin{figure}
\begin{center}
\includegraphics{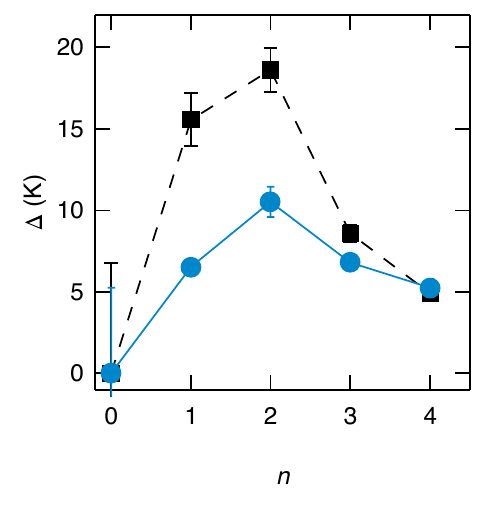}
\vspace{-0.15 in}
\end{center}
\caption{\small{
Square markers represent the gap by integrating $\partial \mu/\partial n$ extracted from the capacitance contribution, the same as in Fig.4(a) in the main text.
Circle markers represent the gap obtained from the temperature dependence of the dissipative contribution, $R \sim exp(-\Delta/2T)$.
The error bars for $n > 0$ stands for the standard deviation of the fitting, and the error bar at $n = 0$ stands for the uncertainty in distinguishing the EC at density balance from at density imbalance or IQH gap.
}}
\label{disp_gap}
\vspace{-0.15 in}
\end{figure}

In our experiment, the resistance at EC states are large, giving rise to finite dissipative contribution.
To obtain the gap value, we obtain the quantum capacitance $C_q$ and resistance $R_s$ by numerically solving \req{maineq1} and \req{maineq2}.
An example of the obtained values $C_q$ and $R_s$ are shown in \rfig{RC}.
The chemical potential jump due to the formation of incompressible states was then found by integrating $\partial \mu/\partial n$ over the peaks in the EC region.
Such gap values are plotted in Fig.4(a) of the main text.
In addition, we fit the temperature dependence of $R_s$ at the EC state by an activation behavior $R_s \sim \exp(-\Delta/2T)$.
Such obtained gaps are plotted in \rfig{disp_gap} as circle markers.
The quantitative differences in the gap values may be due to the contact resistance, which are not accounted for in our analysis.
Nevertheless, they demonstrate the same trend, with a maximum at $n = 2$.

\section{S9. Exciton condensate in another device}

\begin{figure*}
\begin{center}
\includegraphics{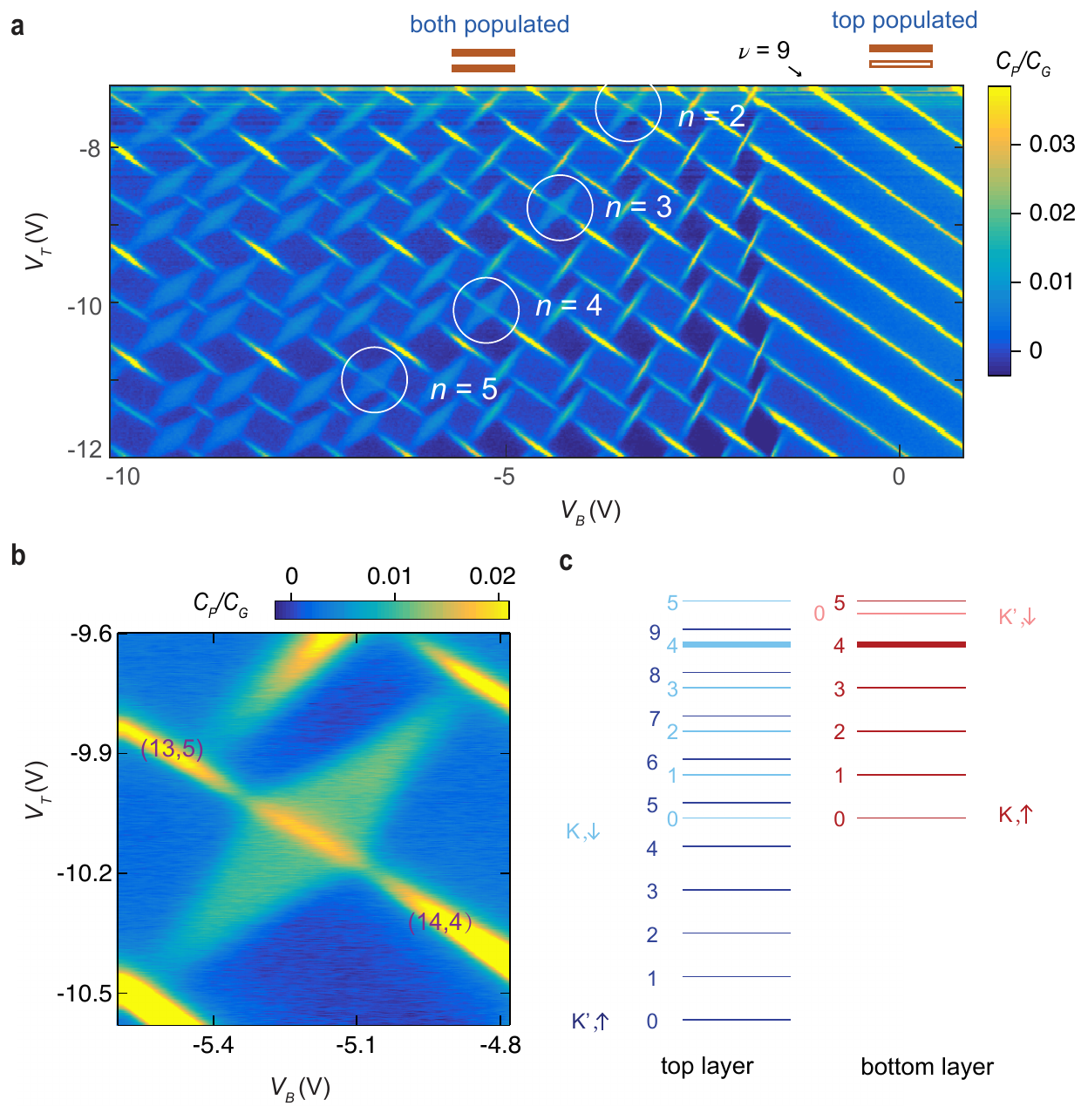}
\vspace{-0.15 in}
\end{center}
\caption{\small{
(a) $C_P/C_G$ as a function of top and bottom gate voltages $V_T$ and $V_B$ at $B$ = 15 T in a different device than in the main text. 
White circles mark the situation where the LL orbital numbers in the top and bottom layers are matched, with different spins.
(b) Zoom-in of (a) focusing on EC with $n = 4$.
The filling factors in adjacent integer quantum Hall states are marked as ($\nu_T$,$\nu_B$).
(c) LL diagram, illustrating the situation of matched orbital number in the two layers. Thick lines mark the relevant LL in panel (b).
}}
\label{ggmap029}
\vspace{-0.15 in}
\end{figure*}


EC with similar behavior as discussed in the main text have been observed in several devices.
Here we show the data in a different device.
In \rfig{ggmap029}(a) we display $C_P/C_G$ versus the top and bottom gate voltages. 
In this device, the top gate also covers the contact region, so $V_T$ is kept at a large negative value in order to highly dope the contact region and maintain good electrical contact.
As a result, the top layer always host a large filling factor. 
For $V_B \gtrsim -1.5$ V, the bottom layer is empty, and only one set of LLs gaps from the top layer are observed;
for $V_B \lesssim -1.5$ V, both layers are populated, and the polarizability features appear, cutting the incompressible features along integer $\nu$ to discontinuous segments.
It can be seen that the layer transitions are sharp when the bottom layer filling $\nu_B$ is low; that is, the carriers in the topmost LL transfers from the top to bottom layer within a small voltage change.
The sharp segments smears out to a diamond shape as $V_B$ and the filling factor difference $\Delta\nu = \nu_T - \nu_B$ decreases.
The white circles mark where the LLs in the two layers host the same orbital index, which corresponds to n = 2, 3, 4 and 5 respectively.
Within these ``diamonds", we observe an incompressible feature when $\nu$ is integer while $\nu_B$ and $\nu_T$ are both fractions.
\rfig{ggmap029}(b) shows a highlight for $n = 4$, with the LL structure sketched in \rfig{ggmap029}(c).
In contrast, the incompressible features are not present in the cases where the LL orbital indices are mismatched across the two layers.

\begin{figure*}
\begin{center}
\includegraphics{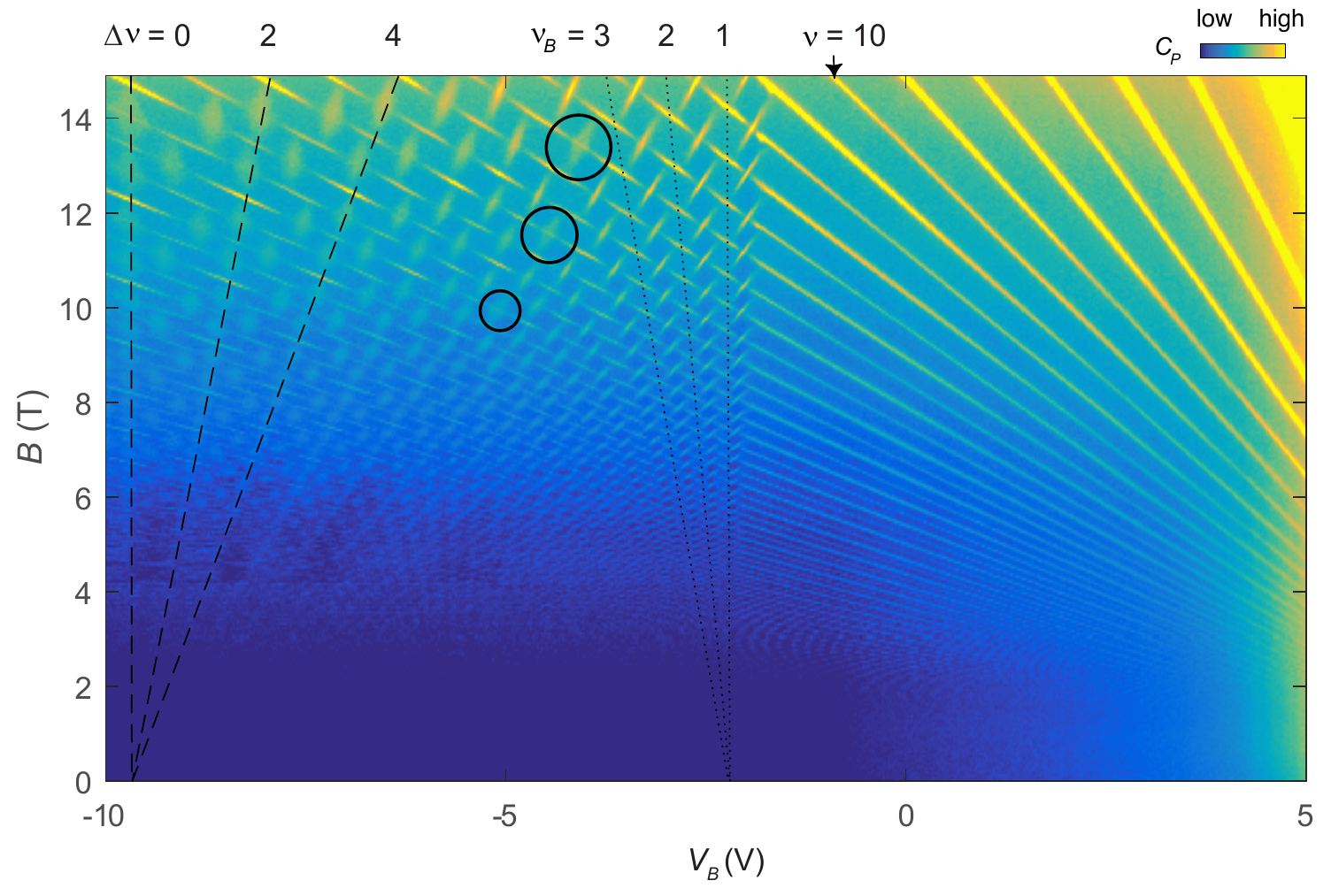}
\vspace{-0.15 in}
\end{center}
\caption{\small{
Penetration signal as a function of magnetic field $B$ and bottom gate voltage $V_B$ at fixed top gate voltage $V_T$ = -8 V. Dotted lines mark the trajectory of $\nu_B$ = 1, 2, 3; dashed lines mark the trajectory of $\Delta\nu = \nu_T - \nu_B$ = 0, 2, 4. Black circles mark the situation where the LL orbital numbers in the top and bottom layers are matched.
}}
\label{fan029}
\vspace{-0.15 in}
\end{figure*}

In addition, in \rfig{fan029} we plot $C_P$ as a function of the bottom gate voltage $V_B$ and $B$ with the top gate voltage $V_T$ fixed.
Three fans of high penetration signal are revealed.
First, a main fan tracks constant integer fillings of $\nu = \nu_T +\nu_B$ and emerges from the band edge. 
For $V_B < 1.5$ V, as both layers are populated, this fan is cut by layer transition features and becomes discontinuous.
Second, a side fan emerges at the population onset of bottom layer, $V_B = 1.5$ V, indicated by dotted lines.
Third, a fan tracks constant $\Delta\nu = \nu_T - \nu_B$, marked by dashed lines.
While the first two fans are both indications incompressible states at integer LL filling,
the third fan is an indication of charge transfer between the layers.
The black circles mark the position where the top most LLs in the two layers host the same LL orbital index, and an incompressible feature appear which signals the emergence of EC.

\section{S10. Theoretical model}\label{sec:numerics}

Our theoretical analysis of bilayer $\mathrm{WSe_2}$ is based on a simplified model that comprises two copies of a two-dimensional electron gas (2DEG), separated by distance $d$ in the perpendicular $z$-direction. The two layers are threaded by the magnetic field $B\hat z$ and they can have arbitrary Landau level (LL) indices, $n_1$ and $n_2$ (for simplicity, we neglect the spin degree of freedom).  The case of $n_1=n_2=0$ and $d/\ell_B \gtrsim 1$ has been studied extensively in previous work on GaAs electron bilayers~\cite{Chakraborty1987, moon:1995, Scarola2001,Schliemann2001, Sheng2003, Park2004, Shibata2006, Moller2009, Papic2015,Zhu2017,Lian2018}. Although the single-particle orbitals in $\mathrm{WSe_2}$ are spinors, the admixture of different LL orbitals is strongly suppressed compared to other materials such as graphene~\cite{rose:2013}. 

In momentum space, our model is described by the Hamiltonian
\begin{eqnarray}\label{eq:ham}
H = \frac{1}{N_\phi}\sum_{\sigma,\sigma'=\uparrow,\downarrow}\sum_{\mathbf{q}} V_{\sigma\sigma'}(\mathbf{q}) \bar\rho_{\sigma}(\mathbf{q}) \bar\rho_{\sigma'}(-\mathbf{q}),
\end{eqnarray}
where $\sigma=\uparrow, \downarrow$ labels the two layers, $N_\phi$ is the number of magnetic flux quanta through the system, $\bar \rho_\sigma$ is the projected electron density in a layer $\sigma$. $V_{\sigma\sigma'}$ is the Fourier transform of the Coulomb interaction which is layer-dependent:
\begin{eqnarray}
V_{\uparrow\uparrow}(\mathbf{q}) &=& V_{\downarrow\downarrow}(\mathbf{q}) = \frac{2\pi }{ |\mathbf{q}|}, \\
V_{\uparrow\downarrow}(\mathbf{q}) &=& \frac{2\pi}{ |\mathbf{q}|} e^{-d|\mathbf{q}|},
\end{eqnarray}
where $d$ is the distance between the layers which ``softens" the repulsion between $\uparrow$ and $\downarrow$ electrons since $V_{\uparrow\downarrow}$ corresponds to interaction $1/\sqrt{r^2+d^2}$ in real space. Thus, $\uparrow$ and $\downarrow$ are sometimes referred to as ``pseudospin" degrees of freedom~\cite{moon:1995} to distinguish them from real electron spin. Furthermore, the effective interaction is also dependent on which LLs the layers are in, which modifies the projected density operators,
\begin{eqnarray}
\bar\rho_\sigma(\mathbf{q}) = e^{-\mathbf{q}^2/4} L_{n_\sigma}\left(\frac{\mathbf{q}^2}{2} \right) \sum_{j=1}^N e^{i\mathbf{q} \mathbf{R}_{j\sigma}}.
\end{eqnarray}
Here $\mathbf{R}_{j\sigma}$ stands for the guiding center coordinate of $j$th electron with the layer index $\sigma$~\cite{PrangeGirvin}, and $L_{n_\sigma}$ is the $n$th Laguerre polynomial for the layer $\sigma$~\cite{Macdonald1994}.

Below we work in units $e^2/\epsilon \ell_B=1$. We focus on the total electron filling factor $\nu=1$, corresponding to the total number of electrons being equal to the number of flux quanta, $N=N_\phi$. If  a layer has LL index $n>0$, we assume that the LLs $0,1,\ldots,n-1$ are fully occupied and inert, while the $n$th LL is partially occupied. We study the model in Eq.~(\ref{eq:ham}) using exact diagonalization of the Hamiltonian matrix appropriate for systems of finite size, either on the surface of a sphere~\cite{Haldane1983} or torus~\cite{Yoshioka1983}. Various symmetries, either spatial (e.g., translation, inversion) or internal  (e.g., pseudospin U(1) symmetry in the absence of interlayer tunneling), are used to block-diagonalize the Hamiltonian and classify its eigenstates. We also interpret our results against Hartree-Fock theory~\cite{Brey1992}, which at $\nu=1$ captures the leading effects as quantum fluctuations are fairly weak given the  small $d/\ell_B$.

\section{S11. Theoretical evidence for exciton condensation}\label{sec:exciton}

Experimental results in the main text demonstrate the quantum Hall effect at integer filling factors when the two layers have matched LL indices, $n_1=n_2$. Here we provide theoretical evidence that identifies this phase as exciton condensate, where each exciton consists of an electron and a hole in the opposite layer (see the recent review~\cite{EisensteinReview}). While the exciton phase exhibits quantized Hall effect, its distinctive signature is the Goldstone mode which governs the gapless excitations of the system in the neutral channel, probed by driving oppositely directed electrical currents through the two layers~\cite{Spielman2000,Tutuc2004}.  

\begin{figure}
\centering
\includegraphics[width=\linewidth]{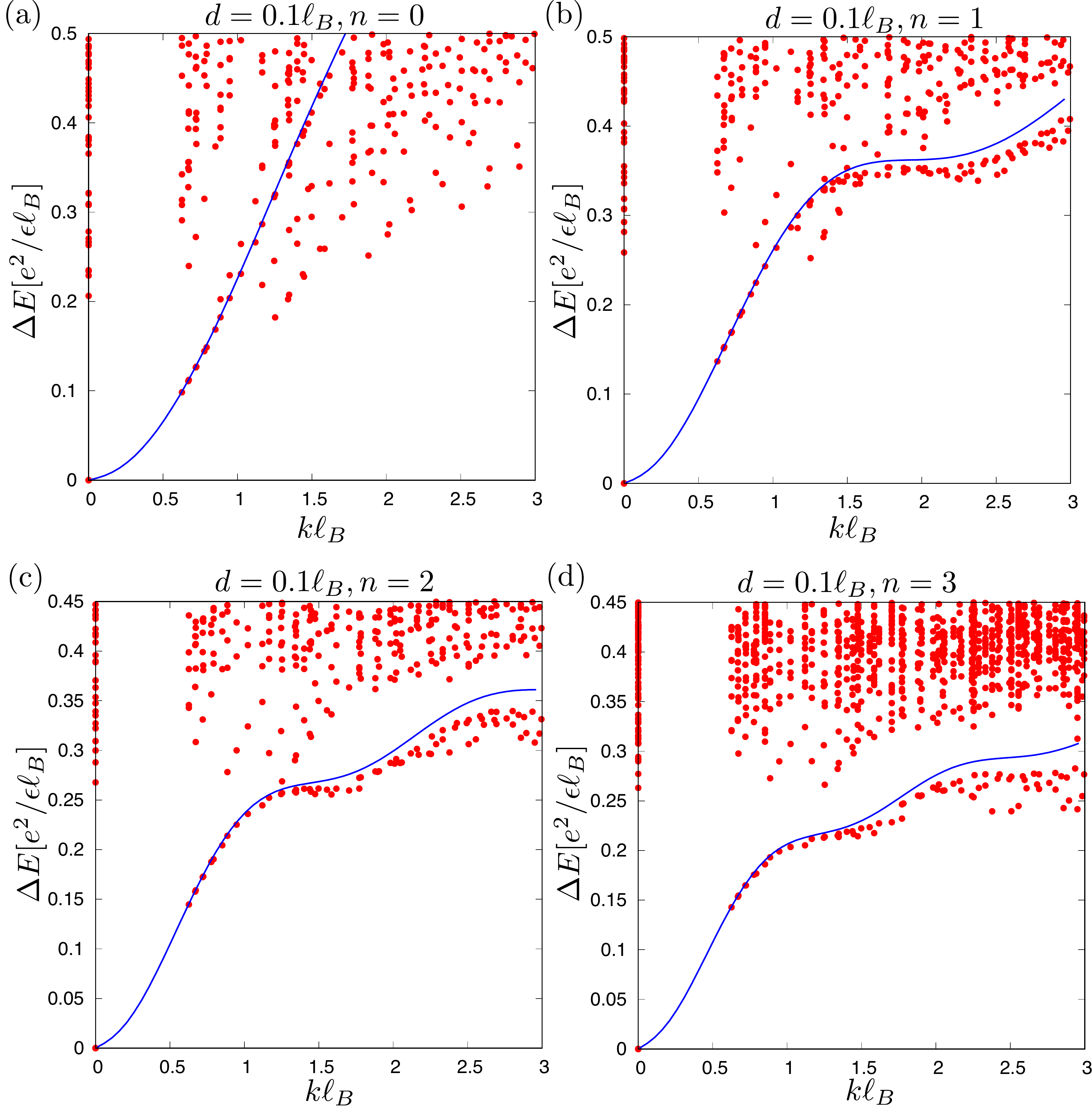}
\caption{Energy dispersion $\Delta E(k)$ as a function of 2D momentum, $k\ell_B$, on a torus geometry. Dots represent exact results for various system sizes ($N=N_\phi=10,12,14,16$ electrons) and boundary conditions (square and hexagon unit cells). In all plots, the two layers have matched LL indices $(n,n)$, where $n=0,1,2,3$ and we fix $d/
\ell_B=0.1$. Lines are the result of the Hartree-Fock theory, which are in excellent agreement with the exact data in the long-wavelength limit. }\label{fig:numerics_goldstone}
\end{figure}

In the numerics we can access the Goldstone mode directly by computing the energy dispersion, $E(k)$, as a function of momentum $k$ in the torus geometry. The torus spectra for a fixed interlayer distance $d/\ell_B=0.1$ and both layers having the same LL index $n=0,1,2,3$ are shown in Fig.~\ref{fig:numerics_goldstone}(a)-(d). We compare exact results against the prediction of Hartree-Fock theory~\cite{Fertig1989}, shown by the blue line in  Fig.~\ref{fig:numerics_goldstone}. We observe, overall, an excellent agreement between the exact results and the Hartree-Fock prediction, in particular in view of the fact that Hartree-Fock results do not contain any adjustable parameters. The agreement could be anticipated given that we are looking at small $d/\ell_B$ and the Hartree-Fock theory captures the ground state exactly for $d/\ell_B=0$. In comparison, much bigger discrepancy between exact results and Hartree-Fock theory are seen already at $d/\ell_B=0.5$, even in the long-wavelength limit (data not shown). 

Results in Fig.~\ref{fig:numerics_goldstone} confirm the existence of a gapless mode with a dispersion in quantitative agreement with the Hartree-Fock theory based on easy-plane XY ferromagnetism~\cite{GirvinMacDonald}. Unfortunately, accessing the long-wavelength of this mode, $k\to 0$, is challenging for exact numerics where the smallest available (non-zero) $k$ scales as $\propto 1/\sqrt{N_\phi}$, thus we cannot resolve finer features of the mode, such as whether the dispersion is quadratic (as expected at $d=0$ when the system is SU(2)-invariant) or linear (for $d>0$). 
Interestingly, the exact spectra in Fig.~\ref{fig:numerics_goldstone} suggest that the Goldstone mode is increasingly better separated from the continuum of the excitation spectrum as LL index is increased, \emph{at all wavelengths} -- see Fig.~\ref{fig:numerics_goldstone}(d). While this  could be due to stronger finite-size effects at larger $n$, the fully exposed collective mode in  bilayer $\mathrm{WSe_2}$ may lead to a different optical response compared to GaAs and graphene materials, where typical $d/\ell_B$ are several times larger and the collective mode is only exposed below the continuum in the range of momenta $k\ell_B \lesssim 1$. 

\begin{figure}
    \centering
    \includegraphics[width=\linewidth]{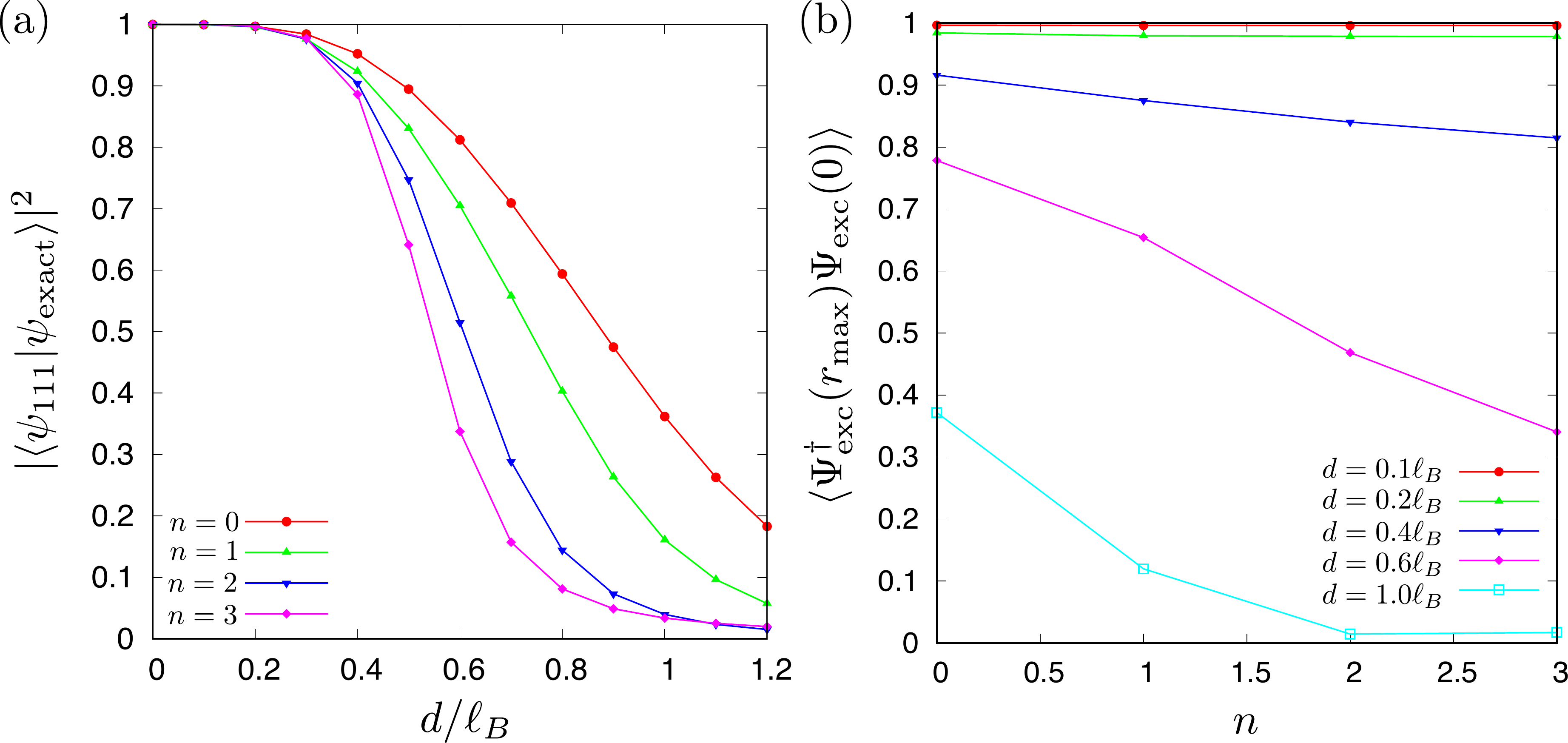}
    \caption{Exciton condensate on the sphere. (a) Overlap between the exact ground state of the bilayer, $\psi_\mathrm{exact}$, and the Halperin 111 state, as a function of interlayer distance $d/\ell_B$ for several matched LL indices $n=0,1,2,3$. (b) Exciton order parameter as a function of matched LL index $n$ and several values of interlayer distance shown in the legend. All data is for $N=14$ electrons on a sphere, where ``0'' denotes the north pole and $r_\mathrm{max}$ is chosen to be the south pole. }
    \label{fig:numerics_exciton}
\end{figure}
Further evidence for the exciton condensate follows from the study of the system's ground state properties. At $d=0$, the exact ground state of the system is given by the first-quantized wave function known as the 111 Halperin state~\cite{Halperin1983}:
\begin{eqnarray}\label{eq:111}
\nonumber \psi_{111}(\{z_i\}, \{ w_j \})& =& \prod_{i<j} (z_i-z_j) \prod_{i<j} (w_i-w_j) \prod_{i,j} (z_i-w_j) \\
&& \times \exp\left(-\sum_k \frac{|z_k|^2}{4\ell_B^2} -\sum_k \frac{|w_k|^2}{4\ell_B^2}\right), \quad
\end{eqnarray}
where $z_i$, $w_i$ are complex 2D coordinates of electrons in top and bottom layers, respectively. The Jastrow factor, $\prod_{i,j} (z_i-w_j)$, builds in the exciton correlation between an electron in one layer ($z_i$) and the corresponding hole at the same position $w_i$ in the opposite layer. Although $\psi_{111}$ is only an exact ground state when strictly $d=0$, it remains a good representative wave function for the entire exciton phase extending to $d\sim \ell_B$~\cite{GirvinMacDonald}. Thus, by evaluating the inner product between the exact ground state of a bilayer and $\psi_{111}$ we can test the existence of the exciton phase in bilayers with $n>0$. These overlaps are shown in Fig.~\ref{fig:numerics_exciton}(a) as a function of $d/\ell_B$ in the sphere geometry. 
We observe that for any value of $n$, as long as $d/\ell_B \lesssim 0.3$, the overlap with 111 state is very close to unity. At larger values of $d$, the overlap drops rapidly, and the decay is faster for larger values of $n$, indicating the sensitivity of the exciton phase in higher LLs to the increase in layer separation.

Further, we do not need to rely on a microscopic candidate wave function to detect the exciton phase. In Fig.~\ref{fig:numerics_exciton}(b) we have computed the exciton order parameter~\cite{Shibata2006}, 
\begin{eqnarray}\label{eq:exciton}
\langle \psi_\mathrm{exact} |\hat  \Psi_\mathrm{exc}^\dagger (r_\mathrm{max}) \hat \Psi_\mathrm{exc}(0)   |\psi_\mathrm{exact}\rangle,
\end{eqnarray}
expressed in terms of an operator that creates an exciton at a given point $\mathbf{r}$,
\begin{eqnarray}
\hat \Psi_\mathrm{exc}(\mathbf{r}) = \hat \Psi_\uparrow(\mathbf{r})^\dagger \hat \Psi_\downarrow(\mathbf{r}),
\end{eqnarray}
where $\hat \Psi_\sigma$ is the standard electron field operator. The normalization of $\hat \Psi_\mathrm{exc}$ is chosen such that the expectation value is 1 when $r\gg \ell_B$ in the ideal exciton case (i.e., for the exact 111 Halperin state). In contrast, a value close to 0 indicates no exciton correlations.  Specifically, on a sphere geometry, we can pick the origin to be the north pole and the point far away is the south pole. The value of the correlator in Eq.~(\ref{eq:exciton}) is shown in Fig.~\ref{fig:numerics_exciton}(b); when $d/\ell_B$ is small ($\lesssim 0.4$), the exciton order parameter is small in any of the LLs ($n=0,1,2,3$). For $d/\ell_B=0.6$ the correlator decays sharply with  $n$, and the exciton phase is only likely to exist in $n=0$ LL. 

\begin{figure}
\centering
\includegraphics[width=\linewidth]{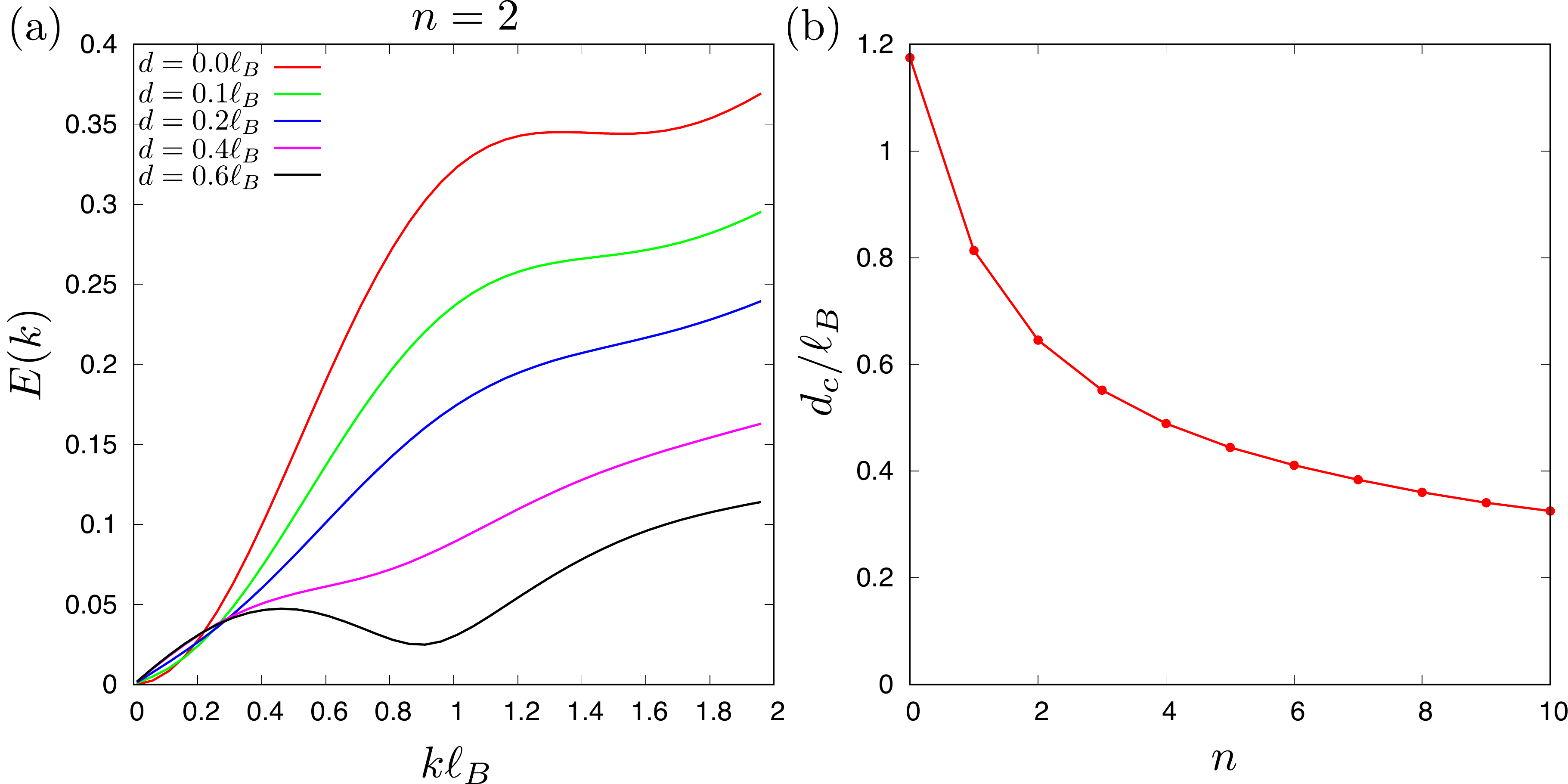}
\caption{(a) A few illustrative examples of exciton dispersion curves $E(k)$ derived in Hartree-Fock theory for $n=2$ and $d/\ell_ B=0, 0.1, 0.2, 0.4, 0.6$. For any $n$, the dispersion develops a minimum that goes soft at a finite $k$ when $d/\ell_B$ is sufficiently large, coinciding with the quantum phase transition to a density wave phase and the destruction of the exciton condensate. (b) Hartree-Fock estimate for the critical distance $d_c/\ell_B$ as a function of LL index $n$. 
}\label{fig:numerics_dc}
\end{figure}
By generalising the Hartree-Fock theory of Ref.~\cite{Fertig1989}, we can conveniently estimate the critical interlayer distance $d_c/\ell_B$ when the exciton condensate is destroyed by a density wave that breaks translation symmetry. In any LL, we find that the exciton dispersion develops a ``roton"-like minimum as $d/\ell_B$ becomes sufficiently large -- see an example of $n=2$ in Fig.~\ref{fig:numerics_dc}(a). The minimum becomes more pronounced by increasing $d/\ell_B$ and eventually goes soft, signalling a quantum phase transition where the ground state changes its momentum from 0 to a finite value $k^* \sim 1/\sqrt{2n+1}$~\cite{Brey2000}. We numerically perform the golden section search and determine the critical value of $d_c/\ell_B$, which is shown in  Fig.~\ref{fig:numerics_dc}(b) as a function of $n$. The critical distance decays sharply for low values of $n$, reaching $d_c/\ell_B \sim 0.3$ for $n\lesssim 10$. Note that this estimate does not include the effect of quantum fluctuations, which become significant near the transition and potentially further decrease $d_c/\ell_B$. Nevertheless, this estimate is consistent with the possibility of the exciton phase in bilayer $\mathrm{WSe_2}$ where we estimate $d/\ell_B \approx 0.1$. We note that in recent works~\cite{shi:2008,zhu:2019exc} the breakdown of the exciton phase was studied at GaAs total filling $\nu=5$, which corresponds to $n=1$ in our case. Refs.~\cite{shi:2008,zhu:2019exc} estimated $d_c/\ell_B \approx 0.9-1.2$, which is somewhat larger than our estimate in Fig.~\ref{fig:numerics_dc}(b), possibly due to finite size effects.

\section{S12. Charged excitations}\label{sec:charge}

\begin{figure}
\centering
\includegraphics[width=\linewidth]{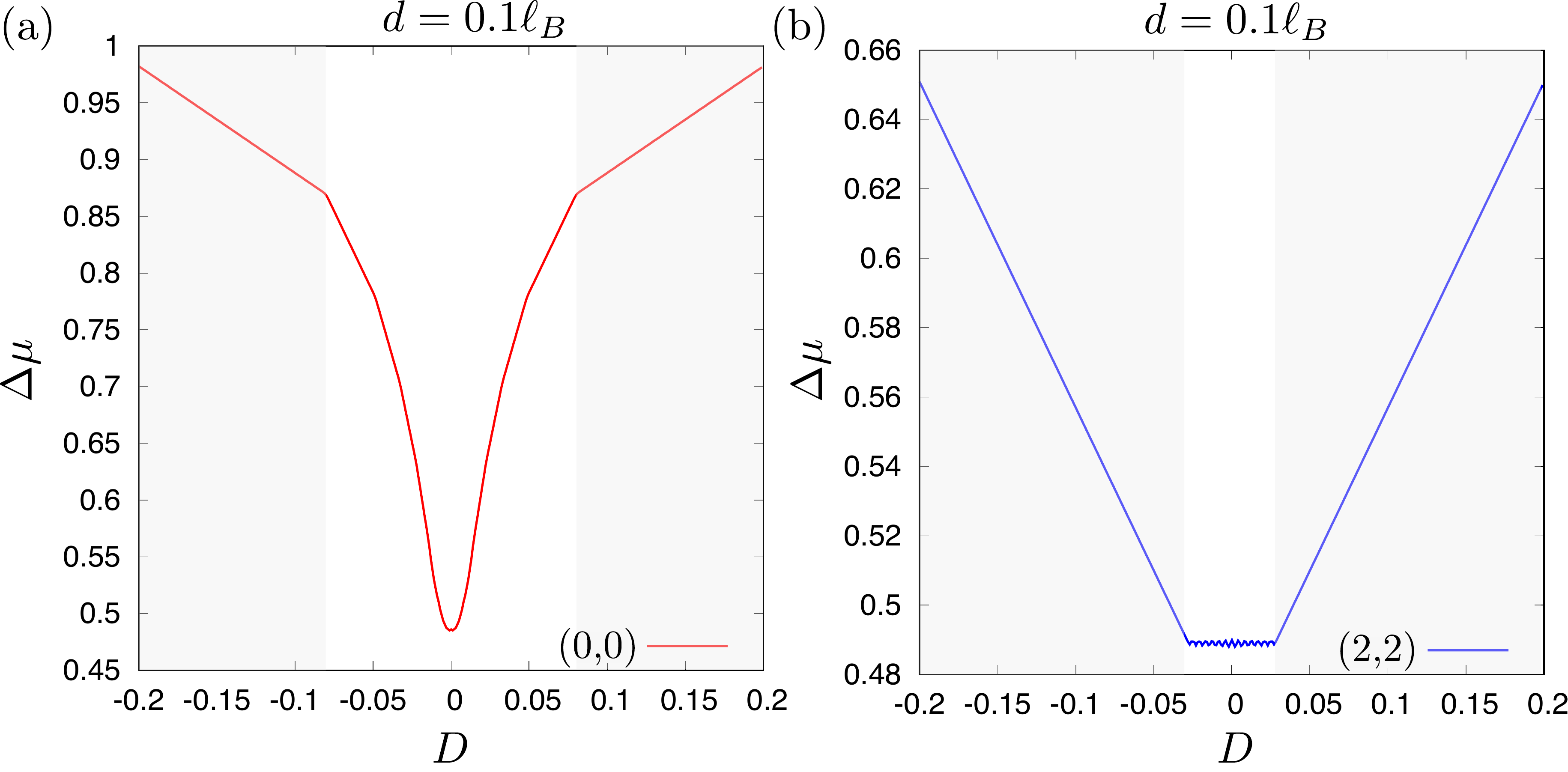}
\caption{Chemical potential discontinuity $\Delta\mu$ (in units $e^2/\epsilon\ell_B$) in Eq.~(\ref{eq:mu}) as a function of charge imbalance $D$, which plays the role of a Zeeman field for pseudospin. (a) $n_1=n_2=0$. (b) $n_1=n_2=2$. Data is for $N=14$ electrons on the surface of a sphere, with $d/\ell_B=0.1$. In shaded regions, the imbalance fully polarizes the pseudospin and the system behaves as a single layer $\nu=1$ state, whose gap varies linearly with $D$. 
}\label{fig:numerics_imbalance}
\end{figure}
We have also studied the response of the exciton phase to adding or removing charge, which is characterized by the chemical potential discontinuity~\cite{Yoshioka1983,moon:1995}:
\begin{eqnarray}\label{eq:mu}
\Delta \mu \equiv E_{N_\phi,N+1} + E_{N_\phi,N-1} - 2E_{N_\phi,N},
\end{eqnarray}
where $E_{N,N_\phi}$ is the ground state energy for the given values of $N$ and $N_\phi$. This quantity plays the role of the activation gap of the system: 
the non-zero value of $\Delta\mu$, in the limit $N\to\infty$, signals an incompressible ground state that would give rise to the quantized Hall effect. Alternatively, one could keep $N$ fixed and vary flux $N_\phi \to N_\phi \pm 1$. At $\nu=1$, it is more convenient to keep the geometry of the system unchanged, i.e., fix $N_\phi$ and vary $N$.

As we are adding or removing charge, it is important to  include the relevant electrostatic corrections when  evaluating $\Delta\mu$ in finite-size systems. For example, the Coulomb interaction gives rise to electrostatic charging energy when particles are shuffled between the layers~\cite{moon:1995}. Formally, this term originates from the finite $\mathbf{q}=0$ contribution in the pseudospin Coulomb interaction, $(V_{\uparrow\uparrow}(0) - V_{\uparrow\downarrow}(0))/2 = \pi d$, and it has a form $\epsilon_c \propto d S_z^2$, where the total pseudospin $S_z \equiv (N_\uparrow - N_\downarrow)/2$. On a finite sphere, the charging energy can be shown to be given by
\begin{eqnarray}\label{eq:ec}
\epsilon_c = \frac{S_z^2}{N^2} \frac{2R^2+1}{R} \left( 1 + \frac{d}{2R} - \sqrt{1+\frac{d^2}{4R^2}}  \right),
\end{eqnarray}
where the radius of the sphere is $R=\sqrt{N_\phi/2}$ (in units $\ell_B=1$).

In Fig.~\ref{fig:numerics_imbalance} we have computed $\Delta\mu$ as a function of density imbalance, $-D(N_\uparrow - N_\downarrow)$. Panel (a) shows the charge gap for $n_1=n_2=0$, while panel (b) illustrates the case of a higher LL with $n_1=n_2=2$. We find the behaviour of the gap of any $n>0$ to be qualitatively similar to each other and distinct from $n=0$ LL. Although we have previously established the ground state for any $n_1=n_2=n$ is the same exciton condensate, approximately described by the 111 state in Eq.~(\ref{eq:111}), we now see that the charged excitations are very different in $n=0$ compared to $n>0$. 

In $n=0$ case, the gap rises sharply upon density imbalance, which is reminiscent of a skyrmion excitation~\cite{sondhi:1993, Schmeller1995} that involves a large number of pseudospin flips. In the absence of Zeeman field ($D=0$), the number of flipped pseudospins is extensive  in system size; for a finite value of $D$, the number of flips is optimized to minimize the total energy due to pseudospin stiffness and the competing Zeeman term. When $D$ is sufficiently large, all pseudospins orient in the $z$-direction [shaded regions in Fig.~\ref{fig:numerics_imbalance}(a)], and the gap then trivially grows linearly with $D$. Adding a small amount of   interlayer tunneling, $-\Delta_\mathrm{SAS} S_x$, does not qualitatively change the behaviour of $\Delta\mu$.  We note that the increase of the gap in $n=0$ GaAs bilayers has been observed experimentally in Refs.~\cite{Spielman2004} and~\cite{Tutuc2003}. In these cases, however, the gap has a much slower parabolic rise with $D$ due to the larger value of $d/\ell_B$. 

In contrast, in higher $n>0$ LLs, the gap is largely insensitive to imbalance until full pseudospin polarization is reached, see Fig.~\ref{fig:numerics_imbalance}(b).  This is consistent with the behavior found in Hartree-Fock approximation which does not include a possibility of pseudospin-texture excitations~\cite{Jungwirth1998, jungwirth:2000}. In the Hartree-Fock picture, the ground state pseudospin smoothly rotates as bias potential is varied. The full pseudospin polarization is reached when the bias potential reaches the value (in the notation of Ref.~\cite{Jungwirth1998})
\begin{eqnarray}
\nonumber U_{\sigma\sigma}^{(n)} &=& \int \frac{d^2 q}{(2\pi)^2} \left( V_{\sigma\sigma} (\mathbf{q}=0) - V_{\sigma\sigma} (\mathbf{q}) \right) \\ && \times  \left( L_n(q^2/2) \right)^2e^{-q^2/2},
\end{eqnarray} 
where $V_{\sigma\sigma}$ denotes the interaction in the pseudospin channel. It follows that $U_{\sigma\sigma}^{(n)}\propto \sqrt{2n+1}$, thus it is harder to polarise the pseudospin in higher LLs, which is consistent with the numerics as well as the experimental data presented in the main text. In Fig.~\ref{fig:numerics_imbalance}(b) tiny wiggles can be observed near $D=0$ -- those correspond to the ground state pseudospin changing in integer steps as a single electron is transferred between the two layers. 

\begin{figure}
\includegraphics[width=\linewidth]{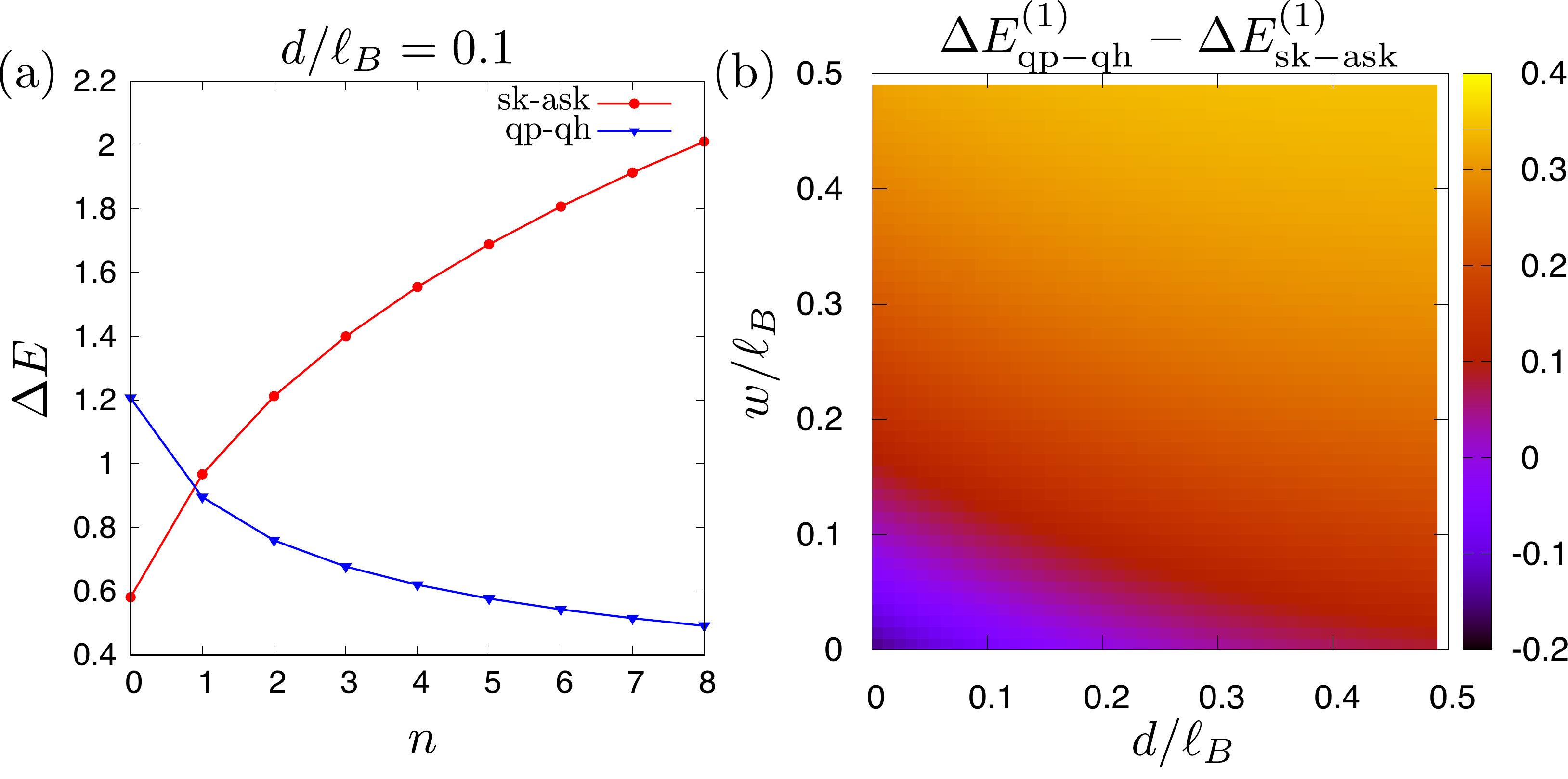}
\caption{(a) Comparison of analytically evaluated gaps for creating a pseudospin texture vs. ordinary electron-hole excitation, as a function of LL index. (b) The difference in energy gap of the pseudospin texture versus the particle-hole (color scale) for $n = 1$, plotted as a function of interlayer distance $d$ and the phenomenological screening parameter $w$ (see text for details). }
\label{fig:numerics_skyrmion}
\end{figure}
The difference in types of charged excitations between $n=0$ and $n>0$ cases is reminiscent of integer quantum Hall ferromagnets, where it has been established that skyrmions are energetically competitive with ordinary electron-hole excitations only in $n=0$ LL~\cite{Wu1994,wu:1995}. In Fig.~\ref{fig:numerics_skyrmion}(a) we analytically compare  the energy gap for creating a spin texture with that for creating an ordinary particle-hole excitation (i.e., with a single pseudospin flipped). Indeed, in $n=0$ we find the skyrmion excitation to be significantly lower in energy~\cite{sondhi:1993}, while in $n=1$ the ordinary particle-hole excitation is already more stable and continues to be energetically favored in all $n\geq 1$. 

Experimental data in the main text shows the gap in bilayer $\mathrm{WSe_2}$ varies non-monotonically with $n$ and reaches a maximum in $n=2$. This suggests that excitations in both $n=0$ (at sufficiently high magnetic fields) and in $n=1$ LL are pseudospin textures. This discrepancy with respect to theory could be explained by disorder, which affects the two types of excitations very differently due to their different spatial profiles, but it could be also be an interaction effect, e.g., due to screening of the Coulomb interaction. In Fig.~\ref{fig:numerics_skyrmion}(b) we have explored the difference in the gap for creating the particle-hole excitation vs. a pseudospin texture in $n=1$ LL, by varying the interlayer distance $d$ and the amount of screening of the interaction. The latter is phenomenologically modelled by finite thickness ansatz, i.e., we model the intralayer Coulomb repulsion as $1/\sqrt{r^2+w^2}$ , while the interlayer potential is $1/\sqrt{r^2+(w+d)^2}$. We observe that either the increase in $d$ or some screening $w \gtrsim 0.1$ is sufficient to stabilize the skyrmion excitation, in line with previous results for the integer quantum Hall ferromagnets~\cite{Cooper1997,Wojs2002}. Thus, it is likely that these interactions effects, perhaps in combination with disorder, stabilize the pseudospin texture in $n=1$ LL of $\mathrm{WSe_2}$ over the particle-hole excitation.

Finally, we comment on the case of \emph{unmatched} LLs, $n_1\neq n_2$. In this case, experimental data in the main text suggest an absence of the exciton phase. In Hartree-Fock theory~\cite{Jungwirth1998,jungwirth:2000}, the system in this case behaves an easy-axis ferromagnet. When bias potential is introduced, the gap varies linearly with $D$ until a critical value when the pseudospin flips, $S_z=N/2 \to -N/2$. The excitation spectrum on the torus is found to be very different from the Goldstone collective mode seen in Fig.~\ref{fig:numerics_goldstone}, possibly due to the existence of domain wall type excitations~\cite{Rezayi2003}.  Similarly, the overlap with the 111 state as well as the exciton order parameter are found to be small and rapidly diminishing with system size.

\end{document}